\documentclass[showpacs
,prd,amsmath,nofootinbib,amssymb]{revtex4}
\usepackage{graphicx}
\usepackage{subfig}
\usepackage{hyperref}
\usepackage{amssymb}
\usepackage[latin1]{inputenc}
\usepackage[english]{babel}
\usepackage{epsfig}
\usepackage{wasysym}
\usepackage{color,xcolor}
\usepackage{pdflscape}
\usepackage{amsmath}
\usepackage{bm}
\usepackage{epsfig}
\usepackage{amsfonts}
\usepackage{dcolumn}

\begin{document}

\title{Regular black holes in $f(G)$ gravity\\}

\author{Marcos V. de S. Silva$^{(1)}$\footnote{E-mail address:marco2s303@gmail.com}, Manuel E. Rodrigues$^{(1,2)}$\footnote{E-mail 
address:esialg@gmail.com}
}

\affiliation{$^{(1)}$Faculdade de Física, Programa de Pós-Graduação em Física, Universidade Federal do Pará, 66075-110, Belém, Pará, Brazil\\
$^{(2)}$Faculdade de Ci\^{e}ncias Exatas e Tecnologia, Universidade Federal do Par\'{a} Campus Universit\'{a}rio de Abaetetuba, 68440-000, Abaetetuba, Par\'{a}, 
Brazil\\
}


\begin{abstract}
In this work, we study the possibility of generalizing solutions of regular black holes with an electric charge, constructed in general relativity, for the $f(G)$ theory, where $G$ is the Gauss-Bonnet invariant. This type of solution arises due to the coupling between gravitational theory and nonlinear electrodynamics. We construct the formalism in terms of a mass function and it results in different gravitational and electromagnetic theories for which mass function. The electric field of these solutions are always regular and the strong energy condition is violated in some region inside the event horizon. For some solutions, we get an analytical form for the $f(G)$ function. Imposing the limit of some constant going to zero in the $f(G)$ function we recovered the linear case, making the general relativity a particular case.
\end{abstract}

\date{\today}

\maketitle



\section{Introduction}
\label{sec1}
Since it was proposed, general relativity has been tested and proved to be quite effective to describe phenomena in the solar system and beyond \cite{din,wal,will1}. Some of the most famous results of general relativity are: the correction of precession of perihelion of mercury \cite{will1,will2}, the bending of light due to the gravitational field \cite{will3} and the existence of gravitational waves \cite{Ligo1,Ligo2,Ligo3,Ligo4}. Another important prediction is the existence of black holes; astrophysical objects whose gravitational interaction is so strong that even light can not escape from \cite{wal,chan}.

Even though it is effective in describing certain phenomena, general relativity presents problems to describe some situations, such as the accelerating expansion of the universe and the galaxy rotation curves \cite{cap}. To solve these problems, we must consider modifications in the field equations. This can be done in two ways. The first is considering changes in the matter/energy sector, through the presence of dark matter and dark energy \cite{massi,michael}. The other alternative would be to modify the geometric sector of the equations, which results in the known alternative theories of gravity \cite{cap2}.

One way to obtain the field equations of general relativity is through the variational principle using the Einstein-Hilbert action \cite{las}. In the case of alternative theories of gravity, we can make modifications in this action in such a way that we obtain new equations of motion \cite{cap2}. In 1980, Starobinsky proposed a model that describes an inflation scenario, where he inserts a term of $R^2$ in the Einstein-Hilbert action \cite{sta}. This kind of modification generates what we call the $f(R)$ theory, where the curvature scalar, in the Einstein-Hilbert action, is replaced by an arbitrary function of this scalar. It is also possible to use $f(R)$ models as substitutes for dark matter and dark energy \cite{cem}. It is possible to modify the action by inserting functions of the stress-energy tensor, $T_{\mu\nu}$, as $f(R,T)$ \cite{harko,jamil,alvarenga} and $ f(R,T,R_{\mu\nu}T^{\mu\nu})$ \cite{sergei1}, where $T$ is the trace of $T_{\mu\nu}$ and $R_{\mu\nu}$ is the Ricci tensor.

The curvature scalar is not the only curvature invariant that can be used to construct the Lagrangian density. There are other invariants like $R^{\mu\nu}R_{\mu\nu}$ and $R^{\mu\nu\alpha\beta}R_{\mu\nu\alpha\beta}$ and even combinations of them as $G=R^2-4R^{\mu\nu}R_{\mu\nu}+R^{\mu\nu\alpha\beta}R_{\mu\nu\alpha\beta}$, Gauss-Bonnet invariant, that could be used. Something interesting about the scalar $G$ is that it is a topological invariant in $3 + 1$ dimensions or lower. So that, if we include the Gauss-Bonnet term in the Einstein-Hilbert action we will not have modifications in the equations of motion. Even if the linear term does not make changes in the equations of motion, it is possible to modify the field equations including nonlinear terms in $G$, which are the $f(G)$ theories \cite{Nojiri1,sergei1,felice,rodrigues5,rodrigues6,sergei2,shamir,sergei3,bamba}. This theory has been used to study the late-time accelerating expansion of the universe \cite{felice}.

Another possibility of an alternative theory of gravity is the Teleparallel Theory (TT), where we consider zero curvature and nonzero torsion of the spacetime \cite{hehl,aldrovandi,maluf}. In this theory, instead of the curvature scalar, we have the presence of the torsion scalar in Lagrangian density and the field equations obtained from the variational principle are equivalent to the Einstein equations. As it has been done in the case of the $f(R)$ theory, it is also possible to generalize the TT theory by replacing the scalar torsion in the Lagrangian density for a general function of $\mathcal{T}$ and with that we obtain the known $f(\mathcal{T})$ theory \cite{harko2,dent,rodrigues7,xu,sergei4,lobo2}, where we can explain the present cosmic accelerating expansion with no dark energy \cite{dent,xu,sergei4}, it's also possible to include the trace of the stress-energy tensor to build a $f(\mathcal{T},T)$ function \cite{lobo3,pace,farrugia,lobo4,ganiou,rodrigues8}. In the context of TT there is also an analogue of the Gauss-Bonnet term, $T_G$, that can be used to construct the $f(\mathcal{T},T_G)$ theory \cite{kofinas,kofinas2,kofinas3}.

In addition to the need for dark matter and dark energy, a problem that arises in general relativity, in the study of black holes, is the presence of singularities \cite{Hawking1,Hawking2,Hawking3,Penrose}, that are points or set of points where the geodesic is interrupted and the physical quantities diverge \cite{Bronnikov}. It is believed that the problem of singularity arises because the theory is classical and that in a quantum theory of gravity this problem would be solved \cite{Hawking1}.

As we do not yet possess this complete theory of quantum gravity, an alternative to avoiding the presence of singularities is the study of regular black holes, which has this name because there are no singularities within it \cite{Stefano}. The first regular solution was proposed by James Bardeen, in 1968 \cite{Bardeen}. The Bardeen metric has a de Sitter core with an equation of state $\rho=-p$, where $\rho$ is the energy density and $p$ is the pressure. This solution also satisfies the weak energy condition and violates de strong energy condition \cite{Beato1,rodrigues4}. Since this solution does not satisfy Einstein equations in vacuum, Ayón-Beato and Garcia have shown that the Bardeen solution can be obtained through Einstein equations with a nonlinear magnetic source \cite{Beato1}. It is also possible to show that this metric is a solution for the case with a nonlinear electrical source, but in this case, it is not possible to obtain a closed form for the electromagnetic Lagrangian \cite{rodrigues4}.

After Bardeen's proposal, several regular solutions were studied in the literature. Most of these were developed in general relativity \cite{Beato2,Beato3,Beato4,Kirill1,Kirill12,Irina,Leonardo2,balart,Nami,Ponce,Wang,Manuel1,Hayward,Culetu1,Culetu2,Fernando2}, others in $f(R)$ theory \cite{Lobo,rodrigues2,rodrigues3} and $f(\mathcal{T})$ theory \cite{rodrigues1}. In Einstein's theory it is also possible to find solutions with rotation \cite{bambi,neves,toshmatov,azreg,DYM,ramon} and even with several horizons \cite{nojiri2}, solutions with multihorizons are also studied in alternative theories of gravity. Regular black holes have already been study in Einstein-Gauss-Bonnet gravity, considering a linear term in the $f(G)$ function, for $4+1$ dimensions \cite{ghosh}.

This work is organized as follows. In the  SEC. \ref{sec2} we use the equations of motion to construct the expressions to the gravity theory and to the electromagnetic, considering a spherically symmetric and static source, and we formulated a theorem for the energy conditions in $f(G)$ gravity. In SEC. \ref{sec3} we first impose the quasi-global coordinate and, through an Ansatz, we rewrite the expressions in terms of a mass function. After that, we show that it is possible to generalize black hole solutions from general relativity to the $f(G)$ theory. In SEC. \ref{sec4} we construct the regular models, where we generalized the solutions already known from general relativity to $f(G)$ gravity. As some expressions are too much large or there is no way to find an analytical form, we analyze the results graphically. Our conclusions and perspectives are present in SEC. \ref{sec5}. We dedicate the Appendix \ref{AF} to show the analytical forms of some functions as $f(G)$ and $L(P)$.

Through this work we consider natural units, where $c=\hbar=G=1$, and metric signature $(+,-,-,-)$.
\section{Nonlinear electrodynamics coupled with $f(G)$ gravity: Equations of motion}
\label{sec2}
The action that describes $f(G)$ gravity coupled with a nonlinear electrodynamics is given by \cite{Nojiri1}
\begin{equation}
S=\int d^4x\sqrt{-g}\left[R+f(G)+4L(F)\right],
\label{action}
\end{equation}
where $L(F)$ is the Lagrangian density of the nonlinear electrodynamics, $f(G)$ represents a general function of the Gauss-Bonnet invariant, $G=R^2-4R^{\mu\nu}R_{\mu\nu}+R^{\mu\nu\alpha\beta}R_{\mu\nu\alpha\beta}$, with $R$ being the curvature scalar, $R^{\mu\nu}$ is the Ricci tensor and $R^{\mu\nu\alpha\beta}$ is the Riemann tensor and $g$ is the determinant of the metric $g_{\mu\nu}$. 

To obtain the field equations, we may vary the action \eqref{action} over $g_{\mu\nu}$ that's results in
\begin{equation}
G_{\mu\nu}+8\left[R_{\mu\alpha\nu\beta}+R_{\alpha\nu}g_{\beta\mu}-R_{\alpha\beta}g_{\nu\mu}-R_{\mu\nu}g_{\beta\alpha}+R_{\mu\beta}g_{\nu\alpha}\right]\nabla^\alpha\nabla^\beta f_G+\left(Gf_G-f\right)g_{\mu\nu}=\kappa^2T_{\mu\nu},\label{fe}
\end{equation}
where the subscript $G$ denotes the derivation with respect to the Gauss-Bonnet invariant, $\kappa^2=8\pi$ and $T_{\mu\nu}$ is the stress-energy tensor defined as \cite{din}
\begin{equation}
T_{\mu\nu}=-\frac{4}{\kappa^2\sqrt{-g}}\frac{\delta \left(\sqrt{-g}L\right)}{\delta\left(g^{\mu\nu}\right)}.
\end{equation}

For nonlinear electrodynamics, $L(F)$ is a general function of the scalar $F=F^{\mu\nu}F_{\mu\nu}/4$, where $F_{\mu\nu}=\partial_\mu A_\nu-\partial_\nu A_\mu$ is the Maxwell-Faraday tensor and $A_\mu$ the gauge potential. The stress-energy tensor for this theory is given by \cite{rodrigues2}
\begin{equation}
T_{\mu\nu}=\frac{1}{4\pi}\left[g_{\mu\nu}L(F)-L_FF_{\mu}^{\ \alpha}F_{\nu\alpha}\right],
\end{equation}
with $L_F\equiv \partial L(F)/\partial F$. If we vary the action \eqref{action} with respect to $A_\mu$, we obtain the modified Maxwell equations,
\begin{equation}
\nabla_\mu \left[F^{\mu\nu}L_F\right]\equiv\partial_\mu\left[\sqrt{-g}F^{\mu\nu}L_F\right].
\label{ME}
\end{equation}
We can always recover the Maxwell theory for $L(F)=F$ and $L_F=1$.

At least, the trace of \eqref{fe} is given by
\begin{equation}
R+8\left[R_{\alpha\beta}+Rg_{\beta\alpha}\right]\nabla^\alpha\nabla^\beta f_G+4\left(f-Gf_G\right)=2\left[FL_F-4L\right].	
\label{te}
\end{equation}

To construct regular solutions, we consider spherically symmetric and static spacetime, which is described by the line element
\begin{equation}
ds^2=e^{a(r)}dt^2-e^{b(r)}dr^2-r^2\left(d\theta^2+\sin^2\theta d\phi^2\right),
\label{ele}
\end{equation}
where $a=a(r)$ and $b=b(r)$ are arbitrary functions of the radial coordinate. In this work we are considering only electrically charged source, so that, we can integrate the modified Maxwell equations \eqref{ME} for the line element \eqref{ele} and we find that the only nonzero component of the Maxwell-Faraday tensor is given by
\begin{equation}
F^{10}(r)=\frac{q}{r^2}e^{-(a+b)/2}L_F^{-1},
\label{EF}
\end{equation}
with $q$ being an integration constant and may be interpreted as the electric charge of the source. By the equation \eqref{EF}, the electric field will be determined since we have $L_F$, which is found solving the field equations.

If we write \eqref{fe} in the mixed form, the nonzero components are
\begin{eqnarray}
&&\frac{e^{-2 b}}{r^2} \bigg\{2 \left(f_G \left(2 a''-3 a'b'+a'^2\right)+6 b' f_G'-4f_G''\right)+e^{b} \left(2 f_G \left(2 a''-a' b'+a'^2\right)-b' \left(r-4f_G'\right)-8 f_G''+1\right)\nonumber\\
&&+e^{2 b} \left(1-r^2 f\right)\bigg\}=2\left[L+\frac{q^2}{r^4}L_F^{-1}\right],\label{eq1}\\
&&\frac{e^{-2 b} }{r^2}\bigg\{\left(e^{b}-1\right) \left(e^{b}-4 f_G a''\right)+a' \left(2\left(e^{b}-3\right) f_G b'+4 \left(e^{b}-3\right) f_G'-re^{b}\right)-2 \left(e^{b}-1\right) f_G a'^2-r^2 e^{2 b}f\bigg\}\nonumber\\
&&=2\left[L+\frac{q^2}{r^4}L_F^{-1}\right],\label{eq2}\\
&&-\frac{e^{-2 b}}{4 r^2} \bigg\{16 r a' f_G''-8 \left(f_G-r f_G'\right) \left(2 a''-3a' b'+a'^2\right)+e^{b} \left(2 \left(8 f_G+r^2\right) a''+\left(a'-b'\right)\left(\left(8 f_G+r^2\right) a'+2 r\right)\right)\nonumber\\
&&+4 r^2 e^{2 b} f\bigg\}=2L.\label{eq3}
\end{eqnarray}
The derivative with respect to the radial coordinate is represented by the prime $(')$. The field equations \eqref{fe} are equivalent to the equations presented in the original work of $f(G)$ gravity (eq. (8) in \cite{Nojiri1}) since we have the same set of equations \eqref{eq1}-\eqref{eq3}. As we are working with $f(G)$ theory, it's also important to calculate the curvature invariants, that are given by
\begin{eqnarray}
&&R=e^{-b}\left[a''+\left(a'-b'\right)\left(\frac{a'}{2}+\frac{2}{r}\right)+\frac{2}{r}\right]-\frac{2}{r^2},\\
&&K\equiv R^{\mu\nu\alpha\beta}R_{\mu\nu\alpha\beta}=\frac{4}{r^4}-\frac{8e^{-b}}{r^4}+\frac{e^{-2 b}}{4 r^2}\bigg[4 r^2 a''^2-2 r^2 a'^3 b'+r^2 a'^4+a'^2 \left(r^2 \left(4a''+b'^2\right)+8\right)\nonumber\\
&&-4 r^2 a' a'' b'+8 b'^2+\frac{16}{r^2}\bigg].
\end{eqnarray}
The Gauss-Bonnet invariant is
\begin{equation}
G(r)=\frac{2e^{-2 b} }{r^2}\left[\left(2a''+a'^2\right) \left(1-e^{b}\right) +\left(e^{b}-3\right) a' b'\right].
\label{G}
\end{equation}
Since we have the equations of motion, we need to solve them to find regular black holes solutions. In general relativity, from the condition $G^{0}_{\ 0}-G^{1}_{\ 1}=0$ in the Einstein tensor, we have that $a+b=0$. In alternatives theories of gravity, it's not necessarily the truth, for example, in the $f(R)$ theory it's is just one of the possibilities \cite{Lobo,rodrigues2,rodrigues3}. If we subtract \eqref{eq1} from \eqref{eq2} we get
\begin{equation}
\frac{e^{-2 b} }{r^2}\left\{\left(12f_G'+e^b\left(r-4f_G'\right)\right)\left(a'+b'\right)+8\left(e^b-1\right)f_G''\right\}=0,
\label{eqfG}
\end{equation}
that is a second-order differential equations in $f_G$. The solution of this equation is given by
\begin{eqnarray}
&&f_G(r)=c_0+\int_1^r \left(c_1 e^{\int_1^{k_3} -\frac{\left(12-4 e^{b(k_1)}\right) \left(a'(k_1)+b'(k_1)\right)}{8 \left(e^{b(k_1)}-1\right)} \, dk_1}\right.-e^{\int_1^{k_3} \frac{\left(4 e^{b(k_1)}-12\right) \left(a'(k_1)+b'(k_1)\right)}{8 \left(e^{b(k_1)}-1\right)} \, dk_1}\times \nonumber\\
&&\left.\int_1^{k_3} \frac{k_2\left(a'(k_2)+b'(k_2)\right) }{8 \left(e^{b(k_2)}-1\right)}e^{b(k_2)+\int_1^{k_2} \frac{\left(12-4 e^{b(k_1)}\right) \left(a'(k_1)+b'(k_1)\right)}{8 \left(e^{b(k_1)}-1\right)} \,
dk_1} \, dk_2\right) \, dk_3.
\end{eqnarray}

We may solve \eqref{eqfG} since we know the relation between the functions $a$ and $b$.  In the next section, we will show a way to generalize solutions constructed in general relativity to $f(G)$ theory.

Since we are interested in studying regular black holes, it is very important to analyze the energy conditions associated with these solutions. In order to analyze the energy conditions, we rewrite \eqref{EF} as the Einstein equations with an effective stress-energy tensor.
\begin{equation}
G_{\mu\nu}=\kappa^2T_{\mu\nu}-8\left[R_{\mu\alpha\nu\beta}-R_{\alpha\nu}g_{\beta\mu}-R_{\alpha\beta}g_{\nu\mu}-R_{\mu\nu}g_{\beta\alpha}+R_{\mu\beta}g_{\nu\alpha}\right]\nabla^\alpha\nabla^\beta f_G-\left(Gf_G-f\right)g_{\mu\nu}=\kappa^2 T^{eff}_{\mu\nu},
\label{Geff}
\end{equation}
where $T_{\mu\nu}^{eff}$ has the components
\begin{equation}
T_{\mu\nu}^{eff}=diag\left(\rho^{eff}(r),-p^{eff}_r(r),-p^{eff}_t(r),-p^{eff}_t(r)\right),
\label{Teff}
\end{equation}
where $\rho^{eff}(r)$, $p^{eff}_r(r)$ and $p^{eff}_t(r)$ are the effective energy density, radial and tangential pressure, respectively. The energy conditions are given by \cite{bamba}
\begin{eqnarray}
SEC(r)&=&\rho^{eff}+p^{eff}_r+2p^{eff}_t \geq 0,\label{EC1}\\
WEC_{1,2}(r)&=&NEC_{1,2}(r)=\rho^{eff}+p^{eff}_{r,t}\geq 0,\label{EC2}\\
WEC_{3}(r)&=&DEC_{1}(r)=\rho^{eff} \geq 0,\label{EC3}\\
DEC_{2,3}(r)&=&\rho^{eff}-p^{eff}_{r,t} \geq0.\label{EC4}
\end{eqnarray}
From \eqref{ele}, \eqref{Geff} and \eqref{Teff} we get
\begin{eqnarray}
&&\rho^{eff}(r)=\frac{e^{-2 b} }{\kappa ^2 r^2}\bigg\{-4 f_G a''+6 b' \left(f_G a'-2 f_G'\right)-2f_G a'^2+2 e^{b} \left(f_G \left(2 a''-a' b'+a'^2\right)+2 b'f_G'-4 f_G''\right)+8 f_G''\nonumber\\
&&+2 r^2 \left(F^{10}\right)^2 L_F e^{a+3 b}+r^2 e^{2b} f+2 r^2 e^{2 b} L\bigg\},\\
&&p^{eff}_r(r)=\frac{e^{-2 b} }{\kappa ^2 r^2}\bigg\{4 f_G a''+2 a' \left(f_G \left(a'-3 b'\right)-6f_G'\right)-2 e^{b} \left(2 f_G a''+a' \left(f_G\left(a'-b'\right)-2 f_G'\right)\right)\nonumber\\
&&-2 r^2 \left(F^{10}\right)^2 L_Fe^{a+3 b}-r^2e^{2 b} f-2 r^2 e^{2 b} L\bigg\},\\
&&p^{eff}_t(r)=\frac{e^{-2 b} }{\kappa ^2 r^2}\bigg\{2 \left(f_G-r f_G'\right) \left(2a''-3 a' b'+a'^2\right)-4 r a' f_G''-2 e^{b} f_G \left(2 a''-a' b'+a'^2\right)-r^2e^{2 b} \left(f+2 L\right)\bigg\}.
\end{eqnarray}

In the way that the equations of motion are written, the effective energy density and the effective pressures are equal to the components of Einstein tensor. In this sense, if we consider the same metric, the energy conditions in general relativity and $f(G)$ gravity will be the same. In \cite{rodrigues3} there is a theorem that says that the energy conditions in $f(R)$ are equal to Einstein theory. We can generalize this theorem to $f(G)$ gravity.

\textbf{Theorem}: Given a solution of \eqref{eq1}-\eqref{eq3}, described by the functions $S_1=\left\{a(r), b(r), f(G), L(F),F^{10}(r)\right\}$, if we have a solution in general relativity described by $S_2=\left\{a(r), b(r), \bar{L}(F),\bar{F}^{10}(r)\right\}$, then the energy conditions are identical for $S_1$ and $S_2$ since $T_{\mu\nu}^{eff}$ in \eqref{Geff} is equal to $T_{\mu\nu}$ in general relativity.

\section{New black hole solutions}\label{sec3}
Imposing the symmetry that we have from the general relativity, $b=-a$, \eqref{eqfG} becomes
\begin{equation}
\frac{8 e^{a} \left(e^{a}-1\right) f_G''}{r^2}=0.
\label{eqfG2}
\end{equation}

In general, $e^a\neq 0$ and $e^a\neq 1$. In this sense, the solution of \eqref{eqfG2} is given by
\begin{equation}
f_G(r)=c_0+c_1r,
\label{fG}
\end{equation}
with $c_0$ and $c_1$ being integration constants. This type of behavior is already known in $f(R)$, where we have the No-Go theorem. To construct the function $f(G)$, we need calculate $G(r)$ from \eqref{G} and invert the function to get $r=r(G)$ to replace in \eqref{fG}. Integrating \eqref{fG} in $G$ we get that
\begin{equation}
f(G)=c_0 G+c_1\int r(G)dG.
\label{f}
\end{equation}

A way to construct regular black holes solutions is introducing a mass function, $M(r)$, through the component $g_{00}$ by
\begin{equation}
e^a=1-\frac{2M(r)}{r},
\label{g00}
\end{equation}
where $M(r)$ satisfies the conditions $\lim\limits_{r\rightarrow 0}\left[M(r)/r\right]\rightarrow 0$ and $\lim\limits_{r\rightarrow \infty}M(r)\rightarrow m$, with $m$ being the ADM mass. In terms of $M(r)$, the Gauss-Bonnet invariant is
\begin{equation}
G(r)=\frac{16}{r^6} \left[r^2 M'^2+r M \left(r M''-4 M'\right)+3 M^2\right].
\end{equation}
Through \eqref{fG} and \eqref{G}, is possible write $f(G)$ in terms of the radial coordinate by
\begin{equation}
f(r)=\int f_G(r)\frac{dG(r)}{dr}dr.
\end{equation}
We may use mass functions that are already known from general relativity and verify if the solutions are black holes or regular black holes. 
From the equations of motion \eqref{eq1}-\eqref{eq3}, the quantities related to the electromagnetic theory are given by
\begin{eqnarray}
&&L=\frac{1}{2 r^6}\left[r^2 M'' \left(16 c_0 M+r^2 (8 c_1+r)\right)-r^6f+16 \left(M-r M'\right) \left(r \left(c_1 r-c_0 M'\right)+3c_0 M\right)\right],\label{Lgeral}\\
&&L_F=2 q^2 r\left\{r^2 \left(2 (16 c_1+r) M'+16c_1 M'^2-r (8 c_1+r) M''\right)+16 c_1 r M \left(r M''-7 M'-2\right)+96 c_1M^2\right\}^{-1},\label{LFgeral}\\
&&F^{10}=\frac{1}{2q r^3}\left\{r^2 \left(-r (8 c_1+r) M''+16 c_1 M'^2+2 (16 c_1+r) M'\right)+16 c_1 r M \left(r M''-7 M'-2\right)+96 c_1 M^2\right\}.\label{F10geral}
\end{eqnarray}
Is also necessary to verify the consistency between the functions $L$ and $L_F$. This is made by
\begin{equation}
L_F=\frac{\partial L}{\partial F}=\frac{\partial L}{\partial r}\left(\frac{\partial F}{\partial r}\right)^{-1},
\end{equation} 
with $F=-e^{a+b}\left(F^{10}\right)^2/2$. To electric sources, it's very difficult to construct an analytical form to $L(F)$, so that is more convenient use another function, that comes from the Legendre transformation \cite{Beato2,Beato3}
\begin{equation}
\mathcal{H}=2 F L_F-L.
\end{equation}
If we define a new tensor $P^{\mu\nu}\equiv L_F F^{\mu\nu}$, is possible show that $\mathcal{H}$ is a function of the scalar $P=P^{\mu\nu}P_{\mu\nu}/4=\left(L_ F\right)^2F$. Using \eqref{EF}, the scalar $P$ is given by
\begin{equation}
P(r)=-\frac{q^2}{2r^4},
\label{P}
\end{equation}
and, in terms of the mass function, $\mathcal{H}$ is
\begin{eqnarray}
&&\mathcal{H}=\frac{1}{2 r^6}\left[16 r M \left(r \left(c_1-(c_0+c_1 r) M''\right)+(4 c_0\right.+7c_1 r) M'\right)-2 r^2 M' \left(8 (c_0+c_1 r) M'+r (8 c_1+r)\right)\nonumber\\
&&\left.-48 M^2 (c_0+2 c_1 r)+r^6 f\right].
\label{H}
\end{eqnarray}
From \eqref{P} is possible invert the function to $r(P)$ and then we substitute in \eqref{H} to find $H(P)$. Using the auxiliary field $P^{\mu\nu}$, we can write $L(P)$ as
\begin{equation}
L(P)=2P\mathcal{H}_P-\mathcal{H},
\end{equation}
with $\mathcal{H}_P=\partial\mathcal{H}/\partial P$.

As the expressions for $f(G)$, $L$ are written in terms of the mass function, each different mass function will generate a different nonlinear electrodynamics and $f(G)$ theory. In \cite{Zerbini} the authors proposed a different way to work with the nonlinear electrodynamics in terms of an auxiliary field $X=q\sqrt{-2F}$ and with that, they could find a closed form to $L(X)$. The methods applied in \cite{Zerbini} and here are equivalent since we must obtain the same energy conditions. 

\subsection{Schwarzschild case}
For the Schwarzschild case, the mass function is the ADM mass, $M(r)=m$, and \eqref{g00} is
\begin{equation}
e^a=e^{-b}=1-\frac{2m}{r}.
\end{equation}
The curvature scalar and the Ricci tensor are zero for Schwarzschild. So that, the Gauss-Bonnet invariant is equal to the Kretschmann scalar.
\begin{equation}
G(r)=\frac{48m^2}{r^6}.
\end{equation}
Since we have $G(r)$, we are able to find the $r(G)$ and construct the function $f(G)$. Then, substituting $r(G)$ in \eqref{f} we get
\begin{equation}
f(G)=c_0 G+\frac{6c_1}{5}\left(48m^2 G^5\right)^{1/6}.
\end{equation}
Some important to emphasize is that, in the Einstein's theory, the Schwarzschild is a vacuum solution, however, in the $f(G)$ theory we have a nonzero stress-energy tensor. If we analyze the components of $T_{\mu\nu}$ we find some kind of anisotropic matter/field, that is very similar to the nonlinear electrodynamics. If we considered the stress-energy tensor for the nonlinear electrodynamics, despite the fact that we don't have charge in the geometry, we will find expressions to the electric field and the nonlinear Lagrangian,
\begin{eqnarray}
L=\frac{16c_1m\left(5r-18m\right)}{10r^5},\\
F^{10}=\frac{32c_1m\left(3m-r\right)}{2qr^3}.
\end{eqnarray}
In the limit of general relativity, $c_1\rightarrow0$, these expressions are zero and then the Schwarzschild solution becomes again a vacuum solution.
\subsection{Reissner-Nordström-anti-de Sitter case}
To work with de Sitter/anti-de Sitter type solutions, we insert a term with the cosmological constant, $\Lambda g_{\mu\nu}$, in the left side of \eqref{fe}. As we are interested in solutions with charge, we will consider the Reissner-Nordström-anti-de Sitter (RNAdS) solution, that is described by the mass function
\begin{equation}
M(r)=m-\frac{q^2}{2r}-\frac{\Lambda r^3}{6},
\end{equation}
where $\Lambda$ is the cosmological constant. The component $g_{00}$ is
\begin{equation}
e^{a(r)}=1-\frac{2m}{r}+\frac{q^2}{r^2}+\frac{\Lambda r^2}{3}.
\end{equation}

\begin{figure}
	\includegraphics[height=5.cm,width=7cm]{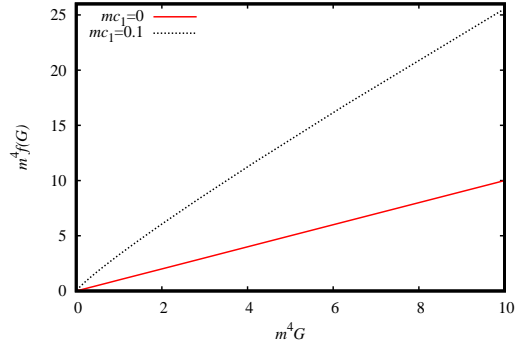}
	\caption{Difference between the behavior of $f(G)$ for $mc_1=0$ and $mc_1=1$, with $q=0.1m$, $m^2\Lambda=0.02$ and $c_0=1$.}
	\label{figfG}
\end{figure}

The curvature and Gauss-Bonnet invariants are
\begin{eqnarray}
&&R(r)=4\Lambda,\\
&&K(r)=\frac{8\left(21q^4-36mq^2r+18m^2r^2+r^8\Lambda^2\right)}{3r^8},\\
&&G(r)=\frac{8\left(15q^4-36mq^2r+18m^2r^2+r^8\Lambda^2\right)}{3r^8}.\label{GRN}
\end{eqnarray}
We can see that this solution presents a singularity at the center of the black hole. From \eqref{GRN} is not possible to construct an analytical form to $r(G)$ and then we cannot find a closed form to $f(G)$. An alternative procedure is calculated $f(G)$ in terms of the  radial coordinate, that is
\begin{equation}
f(G)=\frac{8}{35 r^8} \left(35c_0 \left(6 m^2 r^2-12 m q^2 r+5 q^4\right)+2 c_1 r \left(126 m^2 r^2-245 m q^2 r+100 q^4\right)\right),
\end{equation} 
and with that we can make a parametric plot showing the behavior $f_G(G)\times G$ and $f(G)\times G$. In Fig. \ref{figfG} we show the difference for the linear and nonlinear case of the theory in $G$.

From \eqref{Lgeral}-\eqref{F10geral}, the electromagnetic quantities are
\begin{eqnarray}
&&L(r)=\frac{1}{210  r^7}\left[35 \left(8 c_0 \Lambda ^2 r^7-3 q^2 r^3\right)-8 c_1 \left(35 \Lambda  r^6+600 q^4+42 m r^2 (18 m-5 r)+105 q^2 r (3 r-14 m)\right)\right],\\
&&L_F(r)=6 q^2 r^3\left[6 q^2 r^3+8 c_1 \left(3 q^2 r (5 r-21 m)-r^2 (r-3 m) \left(12 m+\Lambda  r^3\right)+24 q^4\right)\right]^{-1},\\
&&F^{10}(r)=\frac{1}{6 q r^5}\left[6 q^2 r^3+8 c_1 \left(3 q^2 r (5 r-21 m)-r^2 (r-3 m) \left(12 m+\Lambda  r^3\right)+24 q^4\right)\right].
\end{eqnarray}
So that, the coupled between the electromagnetic and gravity theory makes corrections in these functions. If we consider $c_1=0$ and $\Lambda=0$ we recovered the Reissner-Nordström solution in general relativity where the electromagnetic theory is linear with
\begin{equation}
L(F)=F=-\frac{q^2}{2r^4}\ \ \mbox{and}\ \ L_F=1.
\end{equation}

Something interesting to comments is that, if we expand the electric field for points far from the black hole we get
\begin{equation}
F^{10}\approx -\frac{4c_1\Lambda r}{3q}+\frac{4c_1m\Lambda}{q}+\frac{1}{r^2}\left(q-\frac{16c_1m}{q}\right),
\end{equation}
which clearly diverges due to the presence of the cosmological constant. If we have $\Lambda=0$ the electric field is well behaved. In $f(R)$ theory, due to the coupled between the gravity and the electromagnetic theory, the electric field diverges at the infinity of the radial coordinate \cite{rodrigues2,rodrigues3}, that not necessary happen in the $f(G)$ theory. 

As is not possible write an analytical form to $L(F)$, we will show numerically the behavior $L(F)\times F$ and $L(P)\times P$ (the analytical form of $L(P)$ is written in the appendix \ref{AF}). In Fig. \ref{LRN} we show the nonlinear behavior of the electromagnetic theory.

\begin{figure*}
	\includegraphics[height=5.cm,width=7cm]{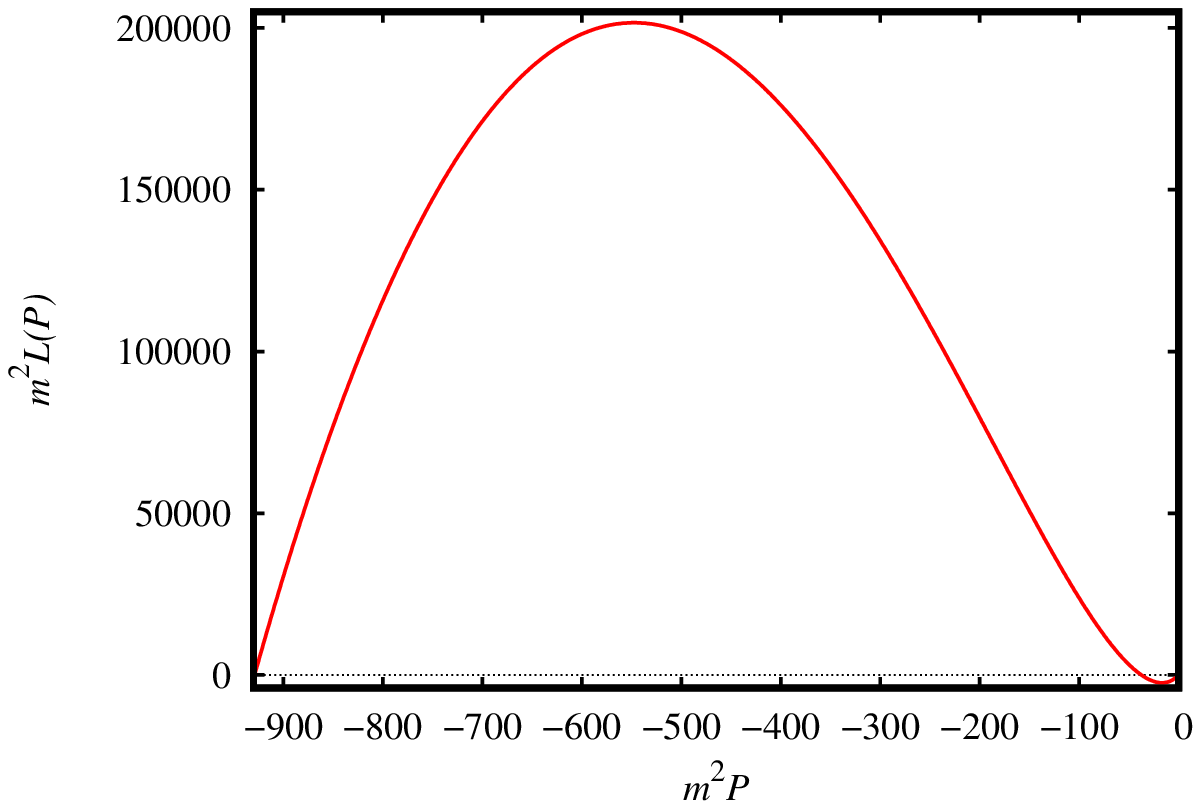}
	\includegraphics[height=5.cm,width=7cm]{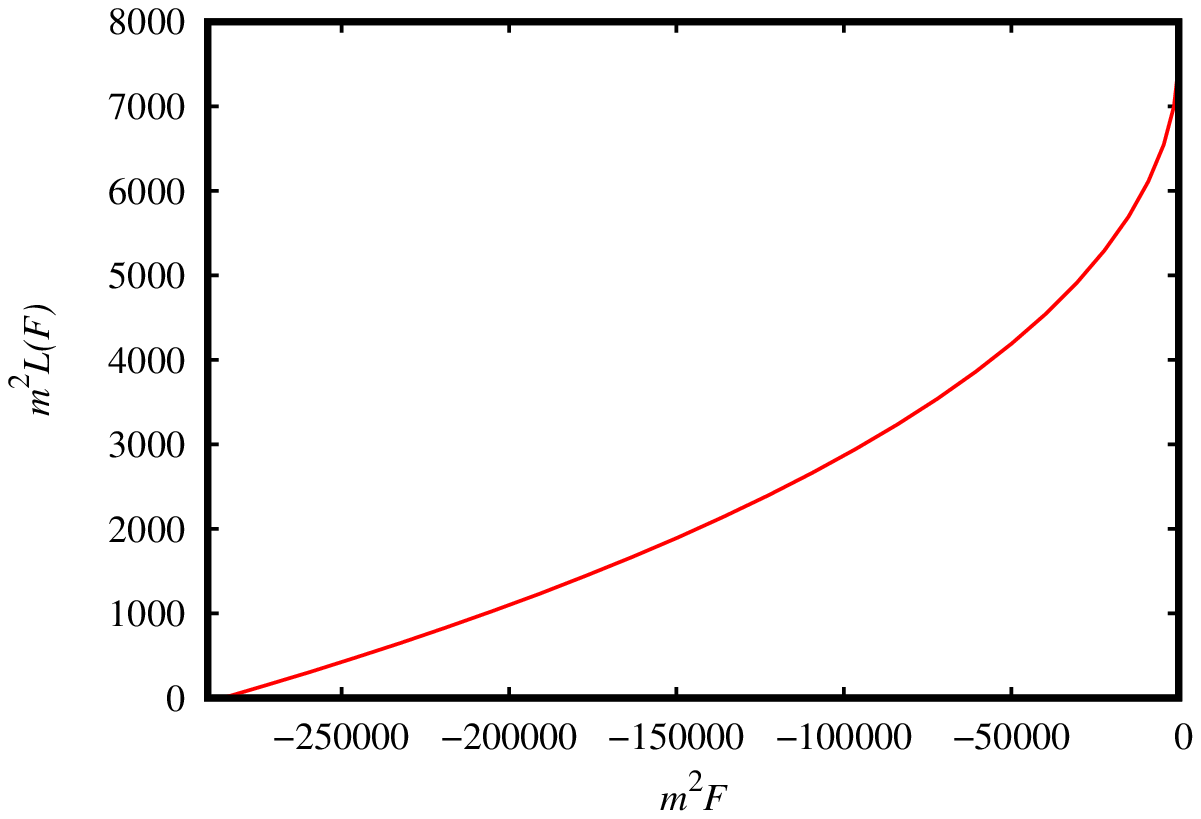}
	\caption{Behavior of $L(F)\times F$ and $L(P)\times P$ for $c_0=1$, $mc_1=1$, $m^2\Lambda=0.02$ and $q=0.4m$.}
	\label{LRN}
\end{figure*}

\section{Regular black hole solutions}\label{sec4}
\subsection{First regular solution}
The first regular solution proposed by Bardeen \cite{Bardeen} may be generated by the mass function
\begin{equation}
M(r)=\frac{mr^3}{\left(r^2+q^2\right)^{3/2}}.
\end{equation}
Some generalizations that have Bardeen as a particular case have already been proposed \cite{balart,Wang}. Let's consider the mass function
\begin{equation}
M(r)=\frac{\alpha m r^{n_1}}{\left(r^{n_2} + q^{n_2}\right)^{n_3}},
\label{genmass}
\end{equation}
where the constant $\alpha$ is a parameter whose unit depends on the constants $n_1$, $n_2$ and $n_3$. The Bardeen solution is recover for $n_1=3$, $n_2=2$ and $n_3=3/2$. To analyze the regular solutions, we will consider the case $n_1=3$, $n_2=2$, $n_3=3/2$ and $\alpha=1$. Than find the following functions
\begin{eqnarray}
&&e^{a}=e^{-b}=1-\frac{2mr^2}{\sqrt{q^6+r^6}},\\
&&K(r)=\frac{12 m^2 }{\left(q^6+r^6\right)^5}\left(8 q^{24}-20 q^{18} r^6+183 q^{12} r^{12}-28 q^6 r^{18}+4 r^{24}\right),\\
&&G(r)=\frac{48 m^2 \left(2 q^{12}-9 q^6 r^6+r^{12}\right)}{\left(q^6+r^6\right)^3}\label{GBDG},\\
&&f(G)=\frac{48 c_0m^2 \left(2 q^{12}-9 q^6 r^6+r^{12}\right)}{\left(q^6+r^6\right)^3}+4c_1m^2\Bigg\{\frac{12 r \left(q^{12}-10 q^6 r^6+r^{12}\right)}{\left(q^6+r^6\right)^3}-\frac{1}{q^5}\Bigg[4 \tan ^{-1}\left(\frac{r}{q}\right)\nonumber\\
&&+\sqrt{3} \ln \left(\frac{q^2+\sqrt{3} q r+r^2}{q^2-\sqrt{3} q r+r^2}\right)-2 \tan ^{-1}\left(\sqrt{3}-\frac{2 r}{q}\right)+2 \tan ^{-1}\left(\frac{2 r}{q}+\sqrt{3}\right)\Bigg]\Bigg\}.\label{fGBDG}
\end{eqnarray}
The analytical forms of $f_G(G)$ and $f(G)$ in terms of the Gauss-Bonnet invariant are given by \eqref{FGGAN1} and \eqref{FGAN1}. With that, we can see that the solution is regular in all points of the spacetime and asymptotically flat. We have the limits $\lim\limits_{r\rightarrow \infty}\left\{e^{a(r)},K(r)\right\}=\left\{1,0\right\}$ and $\lim\limits_{r\rightarrow 0}\left\{e^{a(r)},K(r)\right\}=\left\{1,96m^2/q^6\right\}$. It clearly shows that the solution is regular.

\begin{figure*}
	\includegraphics[height=5.cm,width=7cm]{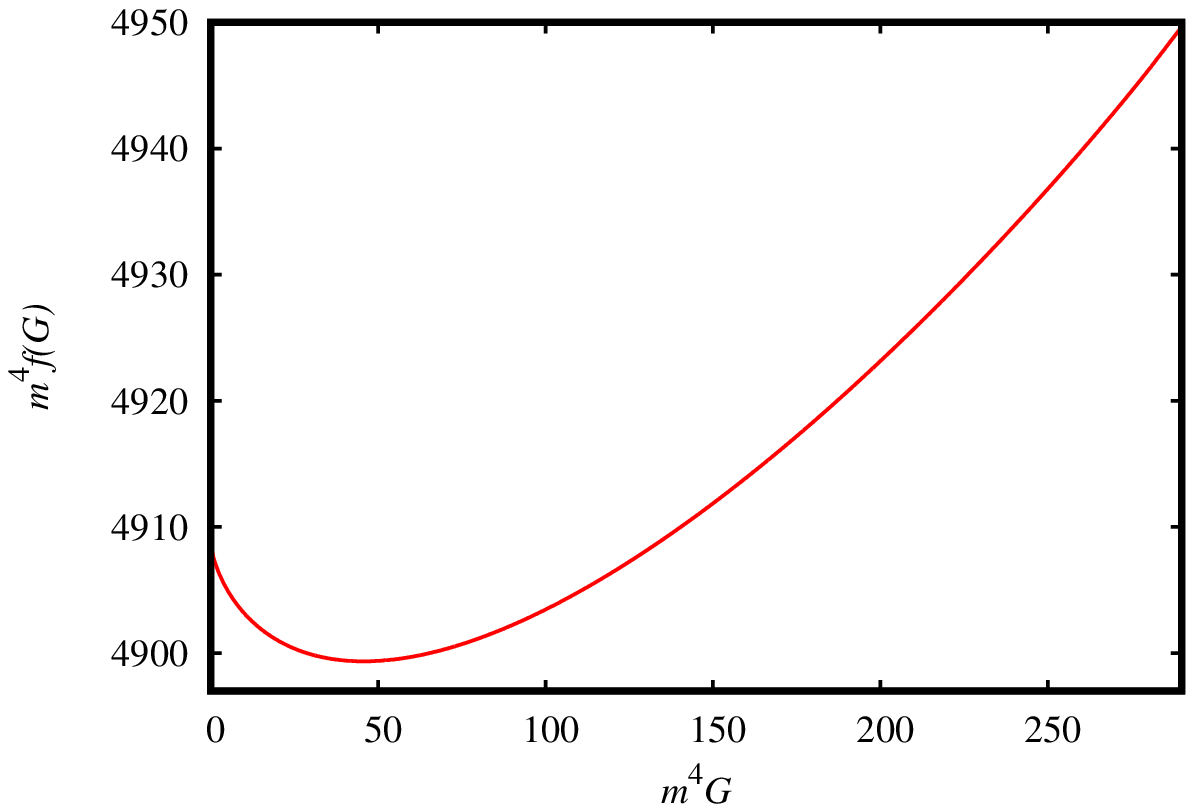}
	\includegraphics[height=5.cm,width=7cm]{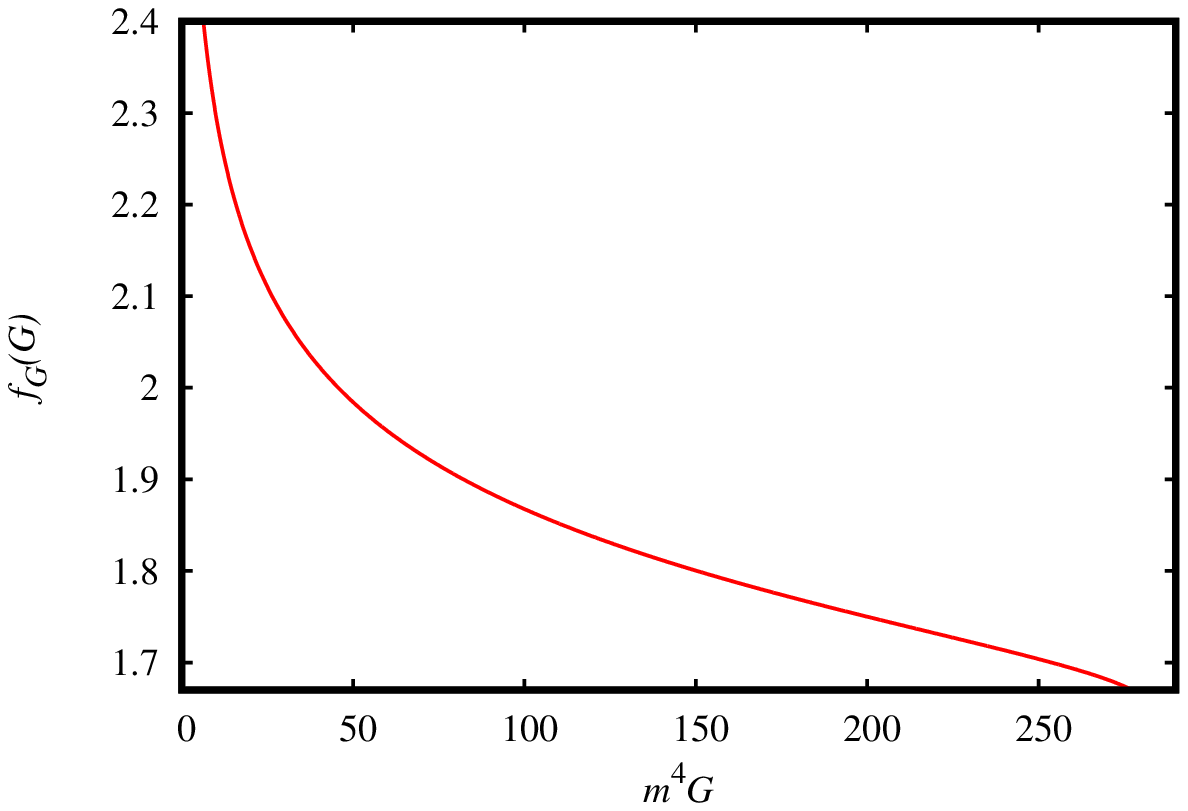}
	\caption{Graphical representation of the functions $f(G)$ and $f_G(G)$ with $c_0=1$, $mc_1=1$ and $q=0.4m$.}
	\label{fBDG}
\end{figure*}

From \eqref{GBDG} we can write $r(G)$ and with \eqref{fG} and \eqref{fGBDG} we construct the functions $f(G)$ and $f_G(G)$, whose nonlinear behavior is shown in the Fig. \ref{fBDG}. We see that $f_G$ diverges for small values of $G(r)$, which corresponds to $r\rightarrow \infty$ in \eqref{fG}. The gravity theory clearly is not general relativity, however, can always be recover to $c_1=0$.

\begin{figure*}
	\includegraphics[height=5.cm,width=7cm]{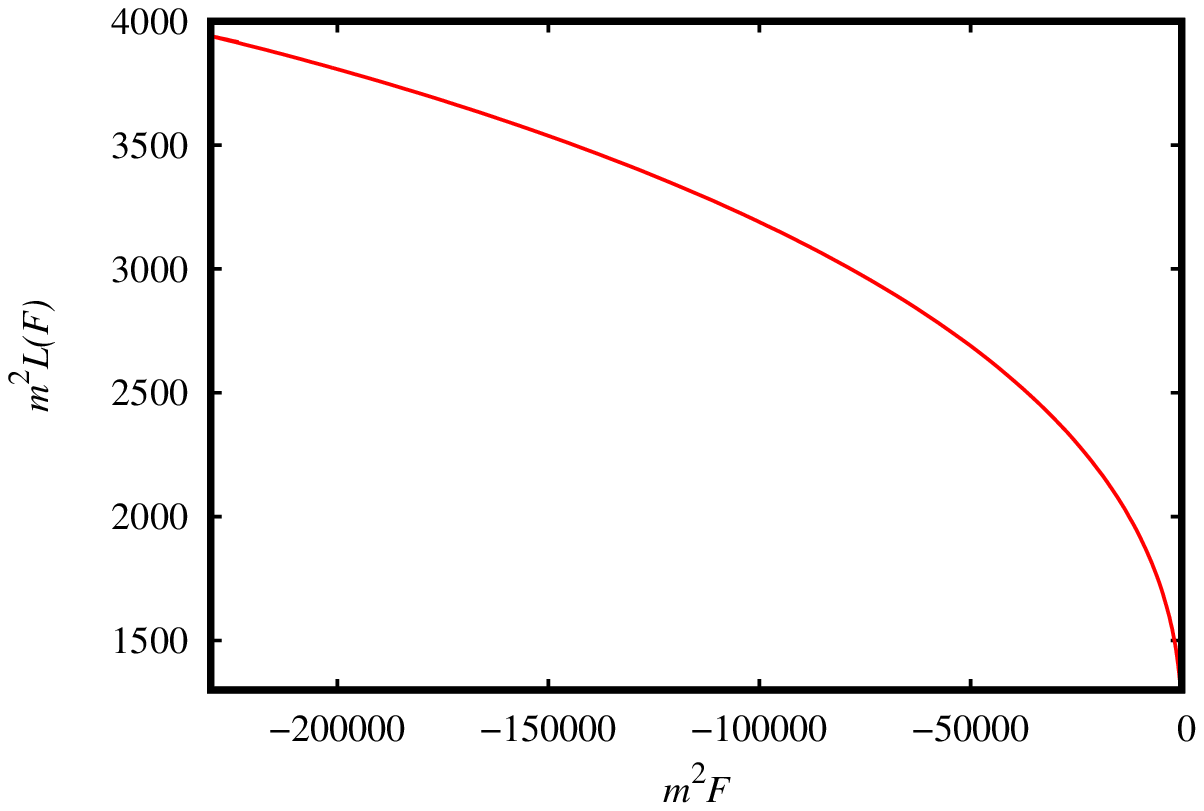}
	\includegraphics[height=5.cm,width=7cm]{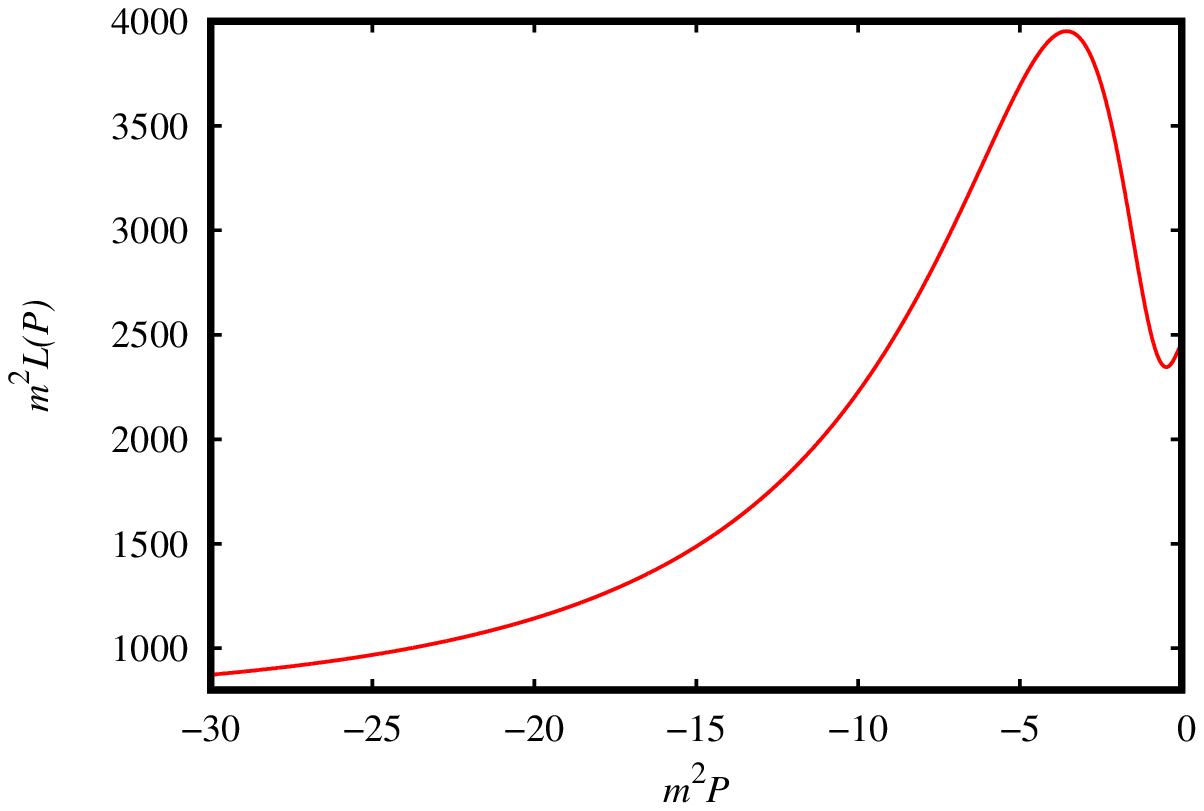}
	\caption{Graphical representation of $L(F)$ and $L(P)$ with $mc_1=1$ and $q=0.4m$.}
	\label{LFPBDG}
\end{figure*}

\begin{figure*}
	\includegraphics[height=5.cm,width=7cm]{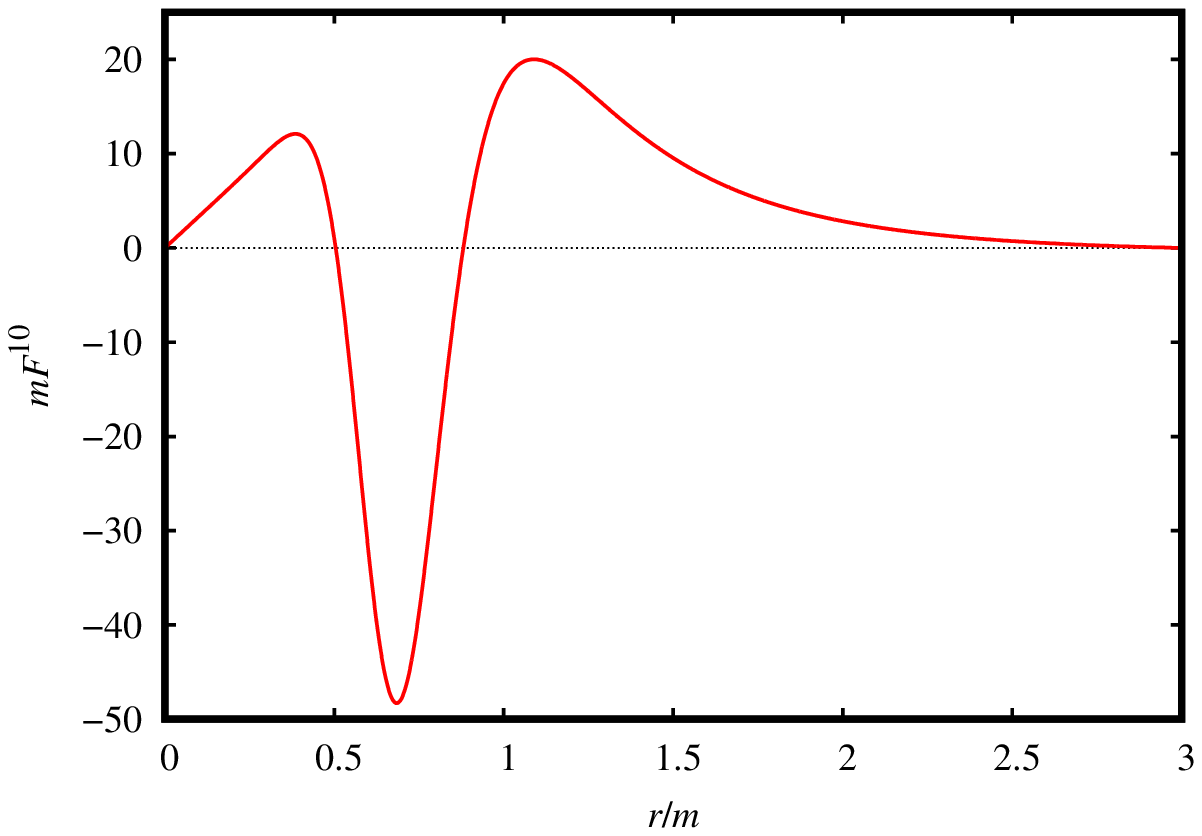}
	\includegraphics[height=5.cm,width=7cm]{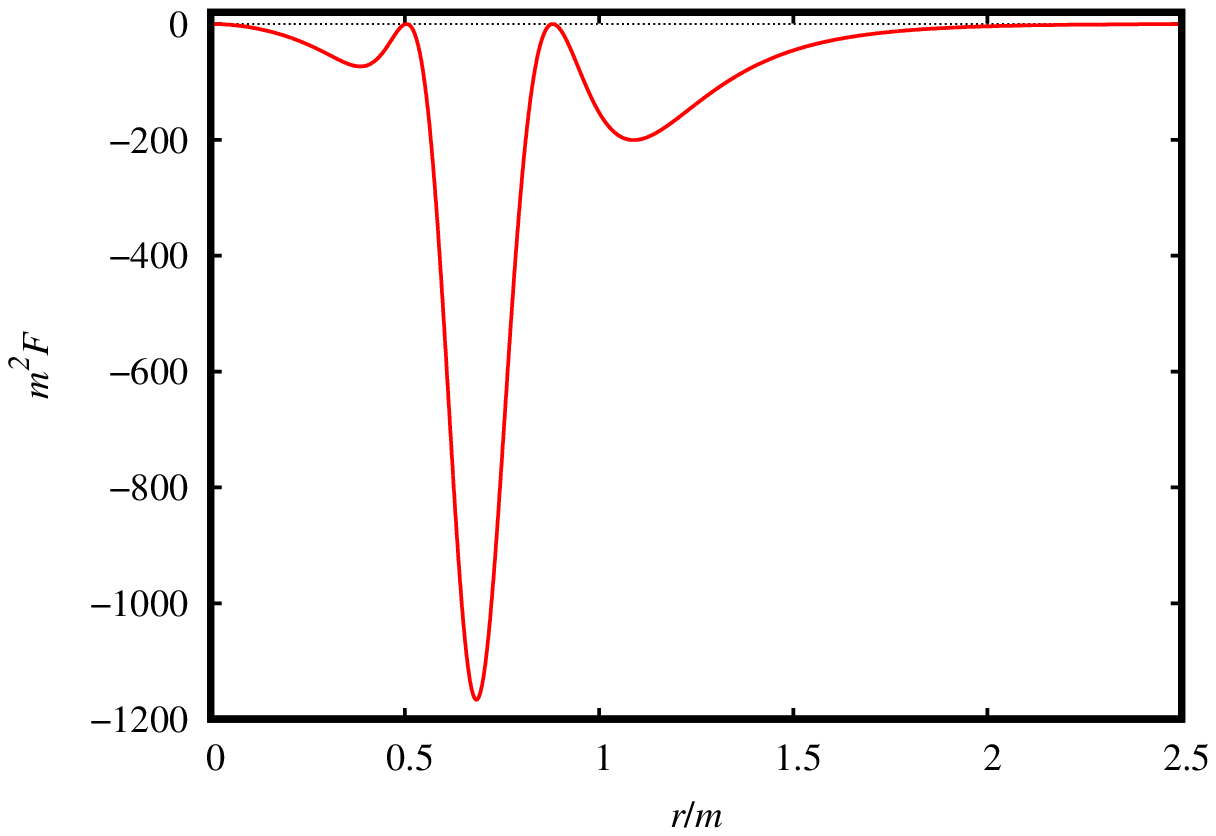}
	\caption{Electric field and the scalar $F$ in terms of the radial coordinate for $mc_1=1$, $\kappa^2=8\pi$ and $q=0.7m$.}
	\label{F0BDG}
\end{figure*}

The next step is calculate the electromagnetic quantities, that are written 
\begin{eqnarray}
&&F^{10}(r)=\frac{m r }{2 q\left(q^6+r^6\right)^3}\bigg\{96 c_1 m r^8 \left(r^6-5 q^6\right)+8 c_1 \sqrt{q^6+r^6} \left(2 q^{12}+25 q^6 r^6-4 r^{12}\right)+27 q^6 r^7 \sqrt{q^6+r^6}\bigg\},\label{F10BDG}\\
&&L(F)=-\frac{24 c_1 m^2 r \left(q^{12}-10 q^6 r^6+r^{12}\right)}{\left(q^6+r^6\right)^3}+\frac{2 c_1 m^2}{q^5} \Bigg(\sqrt{3}\ln \left[\frac{q^2+\sqrt{3} qr+r^2}{q^2-\sqrt{3} qr+r^2} \right]+4 \tan ^{-1}\left(\frac{r}{q}\right)\nonumber\\
&&-2 \tan ^{-1}\left(\sqrt{3}-\frac{2 r}{q}\right)+2 \tan ^{-1}\left(\frac{2r}{q}+\sqrt{3}\right)\Bigg)+\frac{4m c_1 \left(2 q^{12}-23 q^6 r^6+2 r^{12}\right)}{r \left(q^6+r^6\right)^{5/2}}+\frac{3m q^6 \left(2 q^6-7r^6\right)}{2\left(q^6+r^6\right)^{5/2}}\label{L0BDG}.
\end{eqnarray}
Since we have \eqref{P}, \eqref{F10BDG} and \eqref{L0BDG}, we are able to find the behavior of $L(F)$ and $L(P)$. However, is not possible construct $L(F)$ in an analytical form. So that, we show the nonlinearity of the electromagnetic in Fig. \ref{LFPBDG}, which clearly is not Maxwell (the analytical form of $L(P)$ is written in \eqref{LPBDG}). From \ref{F0BDG} we can see some new aspects in relation to general relativity. In the Einstein theory, the electric field and the scalar $F$ of a regular black hole go to zero in the origin and at the infinity of the radial coordinate and the $F^{10}$ has a maximum at the same point where $F$ has a minimum. Here, due to the coupled with the $f(G)$ gravity, the function $F^{10}$ presents more than one extremum, where which extremum corresponds to a minimum in $F$. Expand $F^{10}$ at the infinity we get
\begin{equation}
F^{10}\approx -\frac{16c_1m}{qr^2}+\frac{48c_1m^2}{qr^3}+O\left(\frac{1}{r^4}\right).
\end{equation}
Therefore, the electric field goes to zero at the infinity as in general relativity and different of the $f(R)$ gravity \cite{rodrigues3,rodrigues2}.

\begin{figure}
	\includegraphics[height=5.cm,width=7cm]{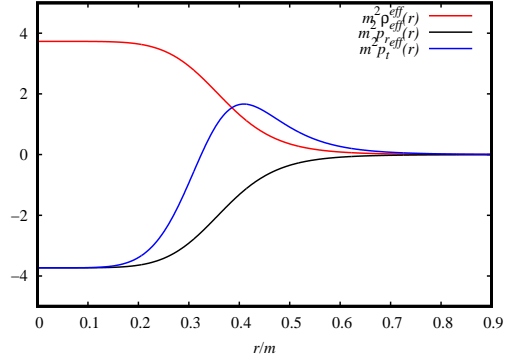}
	\caption{Graphical representation of the fluid quantities for $mc_1=1$ and $q=0.4m$.}
	\label{densBDG}
\end{figure}

Now we will check if the solution satisfies the energy conditions. To do that, we need first calculate the effective energy density and the effective pressures, that are given by
\begin{eqnarray}
\rho^{eff}(r)&=&\frac{6 m q^6}{\kappa ^2 \left(q^6+r^6\right)^{3/2}},\\
p_r^{eff}(r)&=&-\frac{6 m q^6}{\kappa ^2 \left(q^6+r^6\right)^{3/2}},\\
p_t^{eff}(r)&=&\frac{3 m q^6 \left(7 r^6-2 q^6\right)}{\kappa ^2 \left(q^6+r^6\right)^{5/2}}.
\end{eqnarray}
From the fluid quantities, it is possible to notice that the energy density is always positive, that obeys a equation of state of the type $p_r^{eff}=-\rho^{eff}$ and an anisotropic behavior $p_r^{eff}\neq p_t^{eff}$. From the Fig. \ref{densBDG} we can see that close to the center of the black hole we have the behavior of a isotropic fluid $p_r^{eff}\approx-p_t^{eff}$ with a de Sitter equation of state $p^{eff}=-\rho^{eff}$.

Since we have the fluid quantities is easily find the energy conditions. From \eqref{EC1}-\eqref{EC4}, the energy conditions are given by 
\begin{eqnarray}
&&WEC_1(r)=0,\\
&&WEC_2(r)=\frac{27 m q^6 r^6}{\kappa ^2 \left(q^6+r^6\right)^{5/2}},\\
&&DEC_2(r)=2DEC_1(r)=\frac{12 m q^6}{\kappa ^2 \left(q^6+r^6\right)^{3/2}},\\
&&DEC_3(r)=\frac{3 m q^6 \left(4 q^6-5 r^6\right)}{\kappa ^2 \left(q^6+r^6\right)^{5/2}},\\
&&SEC(r)=\frac{6 m q^6 \left(7 r^6-2 q^6\right)}{\kappa ^2 \left(q^6+r^6\right)^{5/2}}.
\end{eqnarray}
The strong energy condition is violated for $r<\sqrt[6]{2/7}\left|q\right|$. As $SEC$ is related with the fact that the gravitational interaction is attractive, inside the regular black hole, we have a surface of zero-gravity, $r=\sqrt[6]{2/7}\left|q\right|$, where inside this surface the gravitational interaction is repulsive. Outside the black hole $DEC_3$ is violated for $r>\sqrt[6]{4/5}\left|q\right|$. This type of behavior is already known from the Bardeen solution \cite{rodrigues4}.

\subsection{Hayward-type solution}
\begin{figure*}
	\includegraphics[height=5.cm,width=7cm]{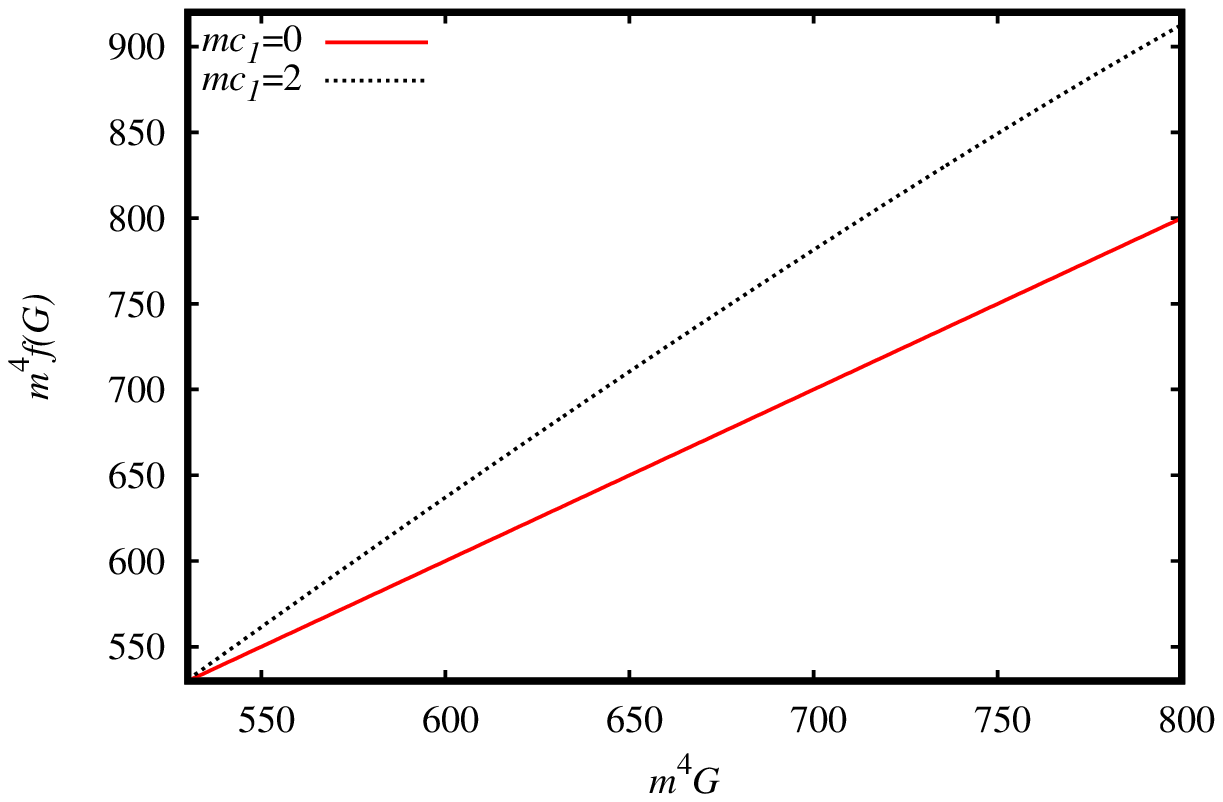}
	\includegraphics[height=5.cm,width=7cm]{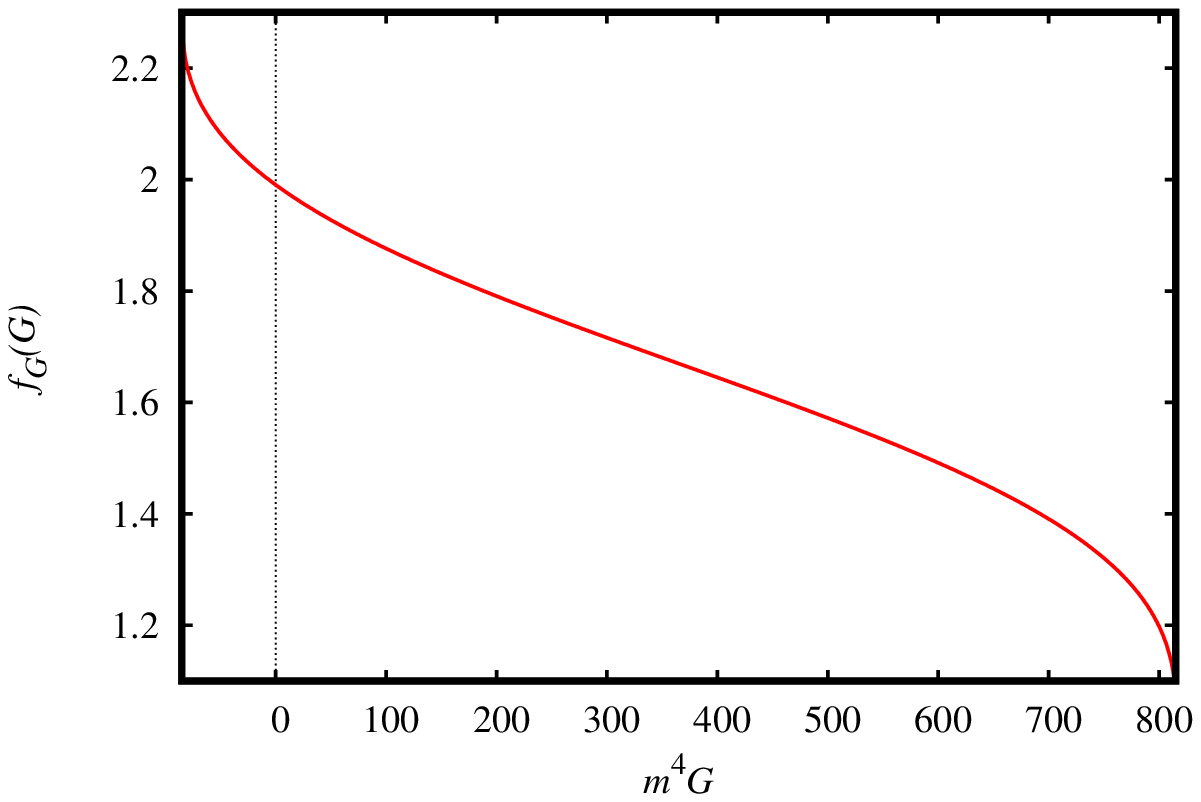}
	\caption{Graphical representation for the  functions $f_G(G)$ and $f(G)$ in terms of the Gauss-Bonnet invariant with $q=0.7m$, $c_0=1$, $mc_1=0,2$ for $f(G)$ and $mc_1=2$ for $f_G(G)$.}
	\label{fGHW}
\end{figure*}

An important regular black hole solution was proposed by Sean Hayward in \cite{Hayward}. If we consider the mass function \eqref{genmass} for $n_1=3$, $n_2=3$, $n_3=1$ and $\alpha=1$ we get a mass function that generates a Hayward-type solution. As we did with the first solution, we get
\begin{eqnarray}
&&e^{a}=e^{-b}=1-\frac{2 m r^2}{q^3+r^3},\\
&&K(r)=\frac{48 m^2 }{\left(q^3+r^3\right)^6}\left(2 q^{12}-2 q^9 r^3+18 q^6 r^6-4 q^3 r^9+r^{12}\right),\\
&&G(r)=\frac{48 m^2 \left(2 q^6-6 q^3 r^3+r^6\right)}{\left(q^3+r^3\right)^4},\label{gHW}\\
&&f(G)=\frac{48 c_0 m^2 \left(2 q^6-6 q^3 r^3+r^6\right)}{\left(q^3+r^3\right)^4}-\frac{16 c_1 m^2}{3 q^5} \left(\frac{3 q^2 r \left(r^9+24 q^6
r^3-2q^9\right)}{\left(q^3+r^3\right)^4}\right.\nonumber\\
&&\left.-\ln \left(\frac{q^2-q r+r^2}{q^2+2qr+r^2}\right)+2 \sqrt{3} \tan ^{-1}\left(\frac{2 r-q}{\sqrt{3} q}\right)\right).\nonumber\\
\label{fHW}
\end{eqnarray}
\begin{figure*}
	\includegraphics[height=5.cm,width=7cm]{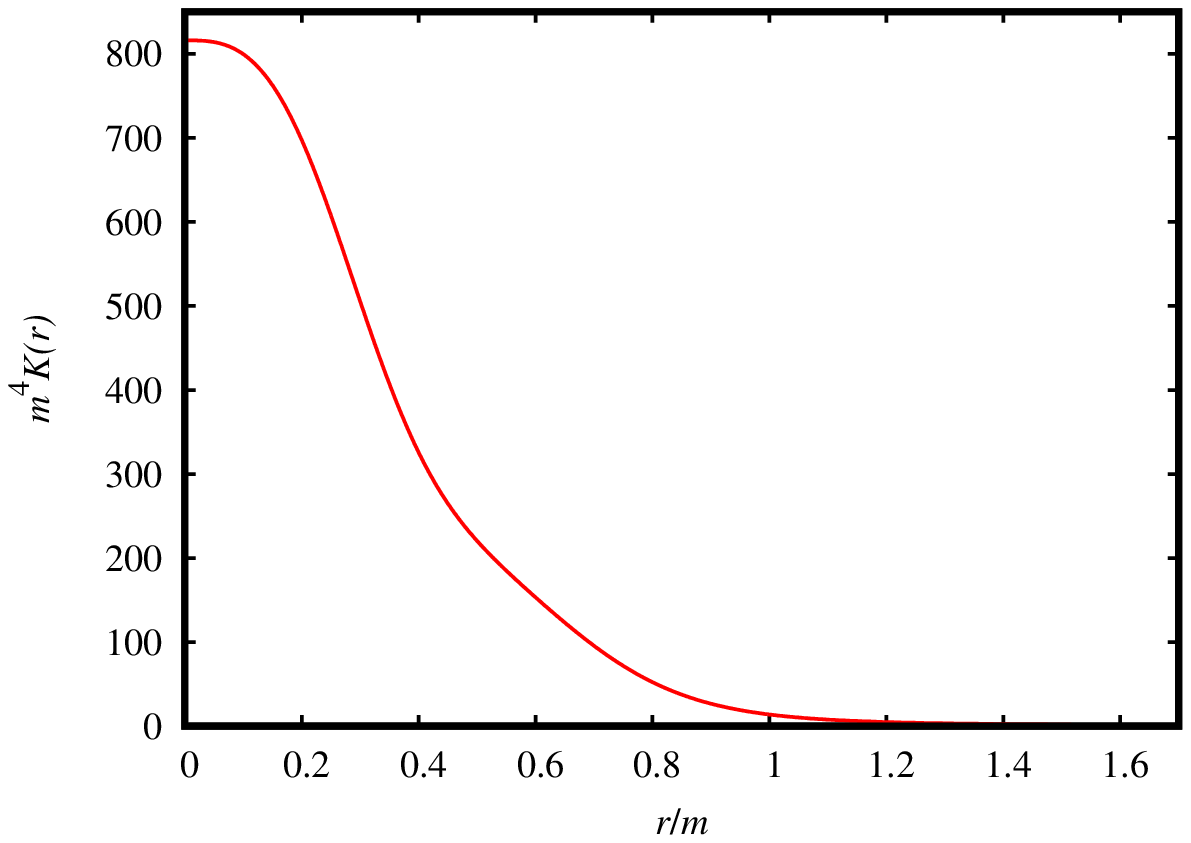}
	\includegraphics[height=5.cm,width=7cm]{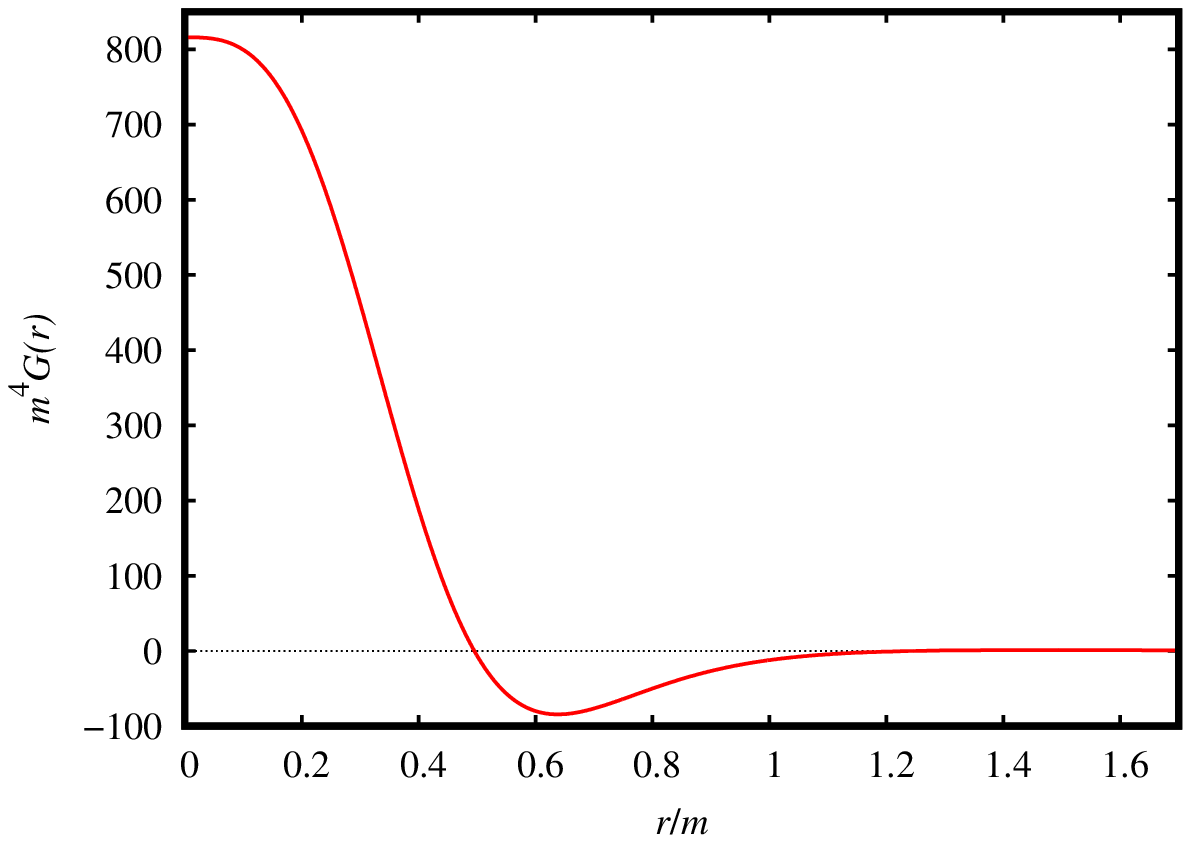}
	\caption{Kretschmann scalar and Gauss-Bonnet invariant in terms of the radial coordinate for $q=0.7m$.}
	\label{GHW}
\end{figure*}
In the Fig. \ref{GHW}, by the behavior of $K(r)$ and $G(r)$, we can see the regularity of the spacetime. For points close to the origin these functions tend to a constant and tend to zero in the infinity of the radial coordinate. From \eqref{gHW} together with \eqref{fG} and \eqref{fHW}, we can prove that the gravity theory is not general relativity. In Fig. \ref{fGHW} we compare the $f(G)$ function for the linear case ($mc_1=0$) with the case that generates the Hayward solution in $f(G)$ gravity (with $mc_1=2$). From $f_G(G)$, it's clearly the nonlinearity of the theory since from the linearity case $f_G(G)$ must be a constant. The analytical expressions are represented by \eqref{anafGHW} and \eqref{anafHW}.

\begin{figure*}
	\includegraphics[height=5.cm,width=7cm]{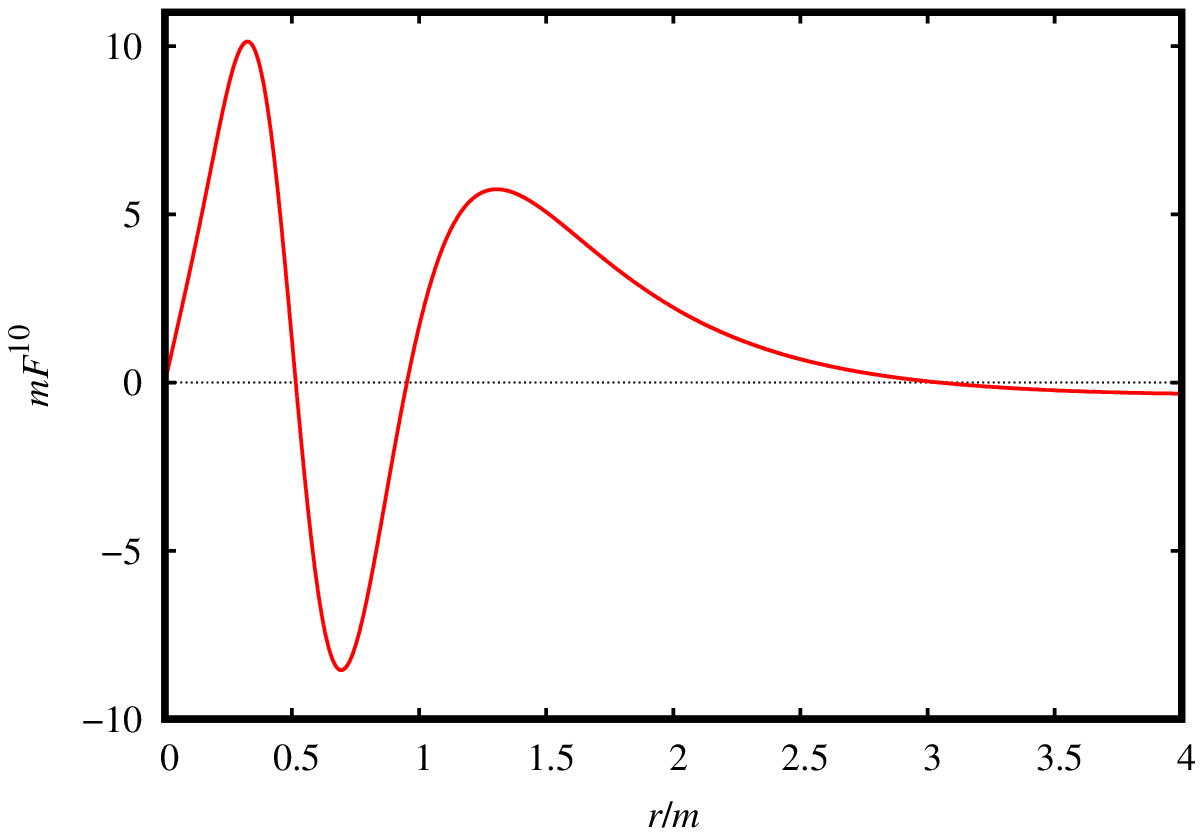}
	\includegraphics[height=5.cm,width=7cm]{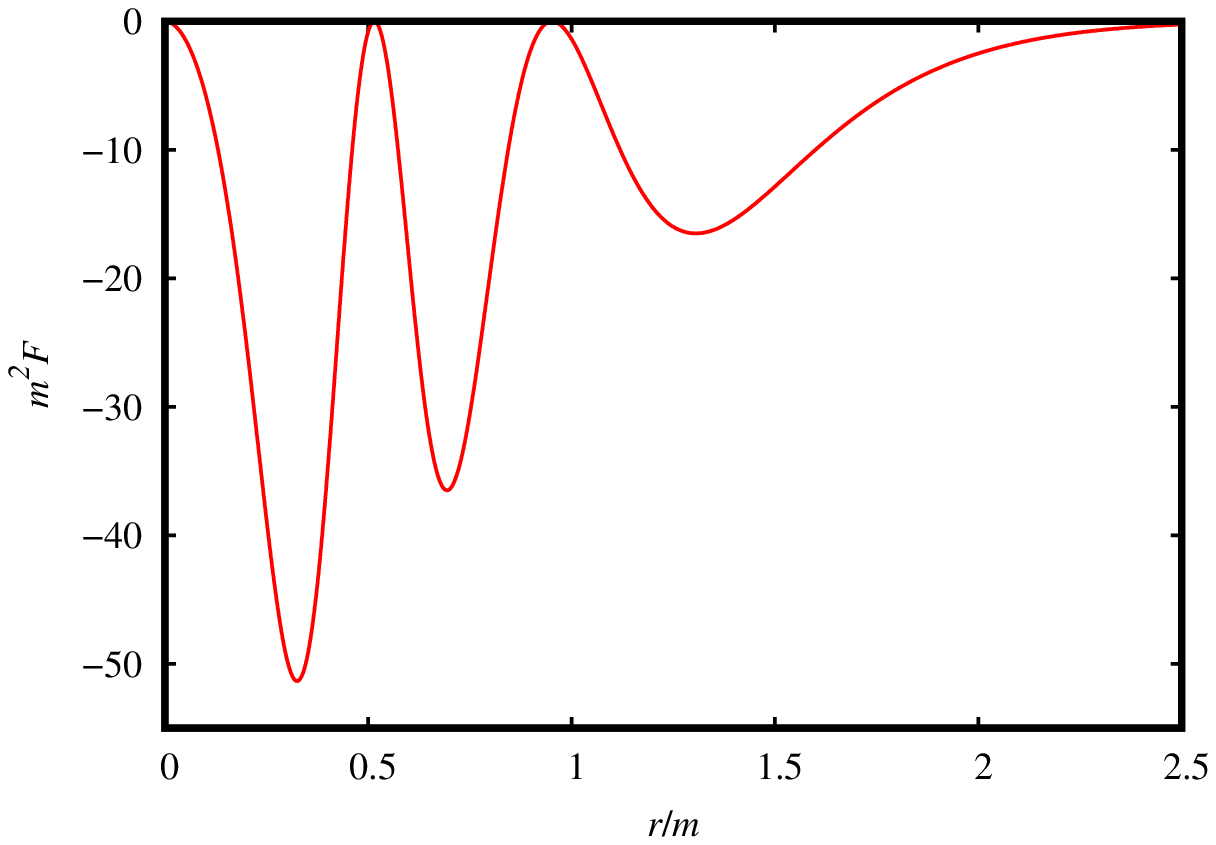}
	\caption{Graphical representation of the electric field and the electromagnetic scalar for $mc_1=1$ and $q=0.7m$.}
	\label{F0HW}
\end{figure*}

\begin{figure*}
	\includegraphics[height=5.cm,width=7cm]{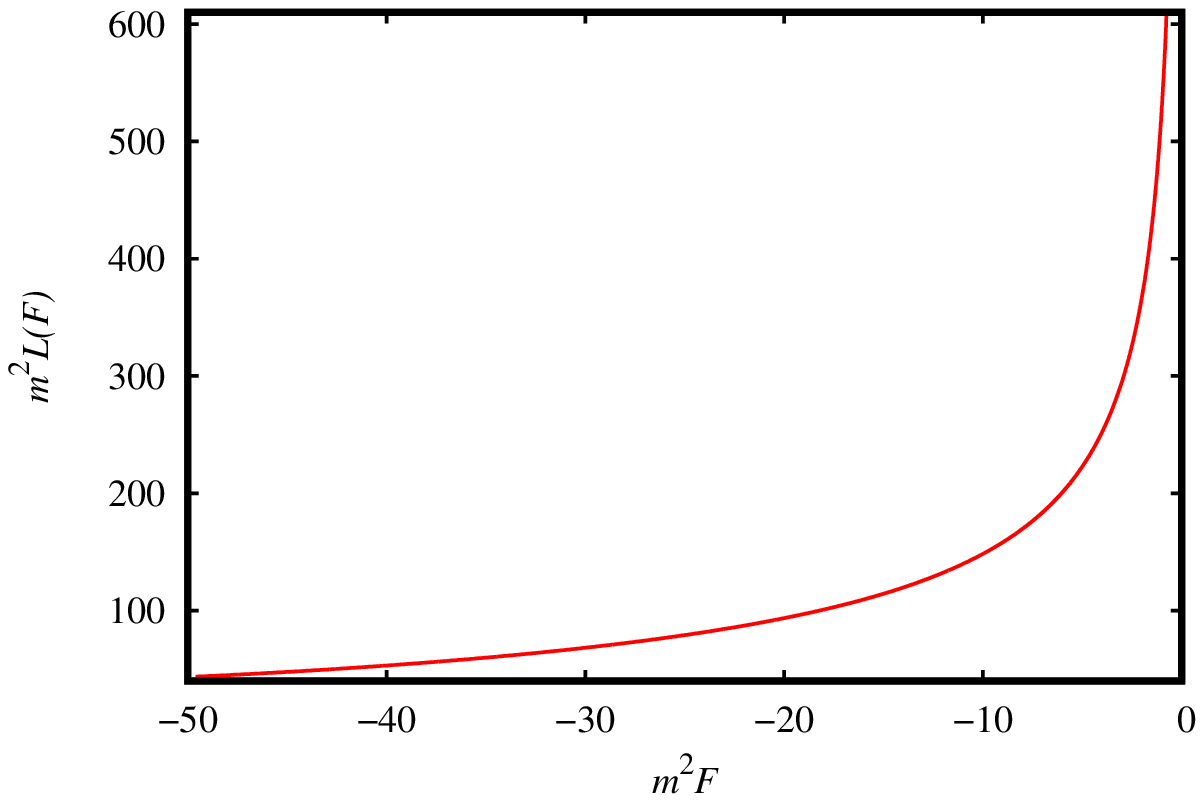}
	\includegraphics[height=5.cm,width=7cm]{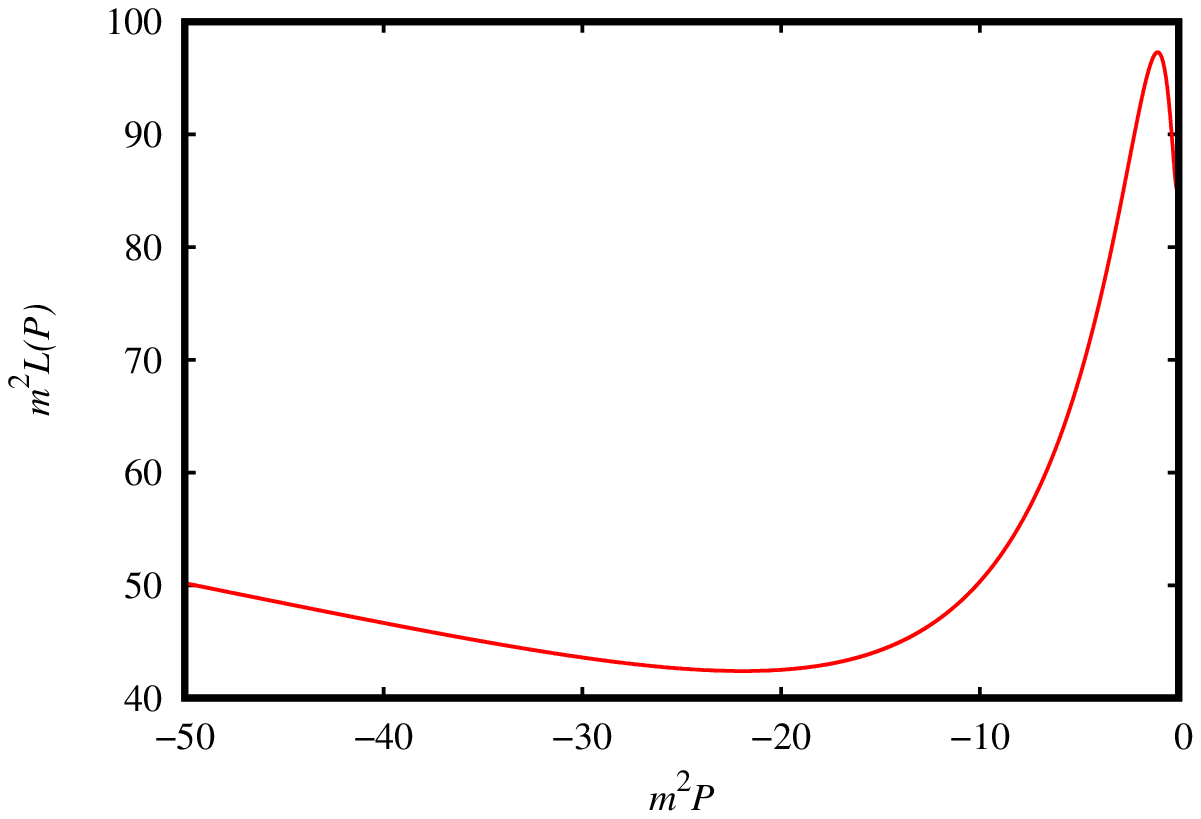}
	\caption{Behavior of the electromagnetic Lagrangian in terms of the electromagnetic scalar and the auxiliary field $P$ for $mc_1=1$ and $q=0.7m$.}
	\label{LHW}
\end{figure*}

From \eqref{Lgeral} and \eqref{F10geral} we have
\begin{eqnarray}
&&L=\frac{2 \sqrt[3]{2m} c_1}{3	q^{10/3}}  \Bigg(\ln \left(\frac{4\sqrt[3]{m2q^4}+4\sqrt[3]{mq^2}r +2\sqrt[3]{2} r^2}{2\sqrt[3]{m^2q^4} - \sqrt[3]{4q^2m} r+\sqrt[3]{2} r^2}\right)+2 \sqrt{3} \tan ^{-1}\left(\sqrt[3]{\frac{4}{mq^2}}\frac{r}{\sqrt{3}}-\frac{1}{\sqrt{3}}\right)\Bigg)\nonumber\\
&&-\frac{m }{3 q^5 r }\Bigg(8 c_1 m r  \left(\ln \left(\frac{q^2-q r+r^2}{(q+r)^2}\right)+2 \sqrt{3} \tan ^{-1}\left(\frac{q-2r}{\sqrt{3} q}\right)\right)\nonumber\\
&&- \frac{24 c_1q^2}{\left(q^3+r^3\right)^4}\left(6 q^6 r^5 (4 m-r)-2 q^9 r^2 (m+3 r)+m r^{11}+q^{12}+q^3 r^9\right)-9 q^8 r \left(q^6-q^3 r^3-2r^6\right)\Bigg),\label{L0HW}\\
&&F^{10}=\frac{m r }{q \left(q^3+r^3\right)^4}\left(8 c_1 \left(3 q^3 r^5 (2 r-7 m)+2 r^8 (3 m-r)+q^9+9 q^6 r^3\right)+9 q^3 r^4 \left(q^3+r^3\right)\right).
\end{eqnarray}
With $F^{10}$ is simple obtain $F$ and then we show the behavior of these functions in Fig. \ref{F0HW}. We can see that the electric field and the electromagnetic scalar are always regular and have zero value in the origin and in the infinity of the radial coordinate. Since we have $F$ and $P$ we may construct $L(F)$ and $L(P)$, whose the nonlinear behaviors are shown in Fig. \ref{LHW}. Finally, the analytical form of $L(P)$ is given by the equation \eqref{LPHW}.

\begin{figure}
	\includegraphics[height=5.cm,width=7cm]{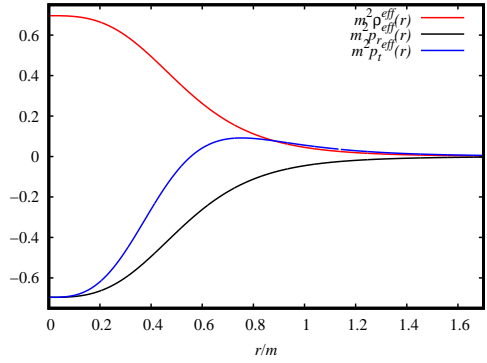}
	\caption{Effective energy density and effective pressures for the Hayward solution for $\kappa=\sqrt{8\pi}$ and $q=0.7m$.}
	\label{densHW}
\end{figure}

At least, in order to analyze the energy conditions, we calculate the effective fluid quantities
\begin{eqnarray}
\rho^{eff}(r)&=&\frac{6 m q^3}{\kappa ^2 \left(q^3+r^3\right)^2},\label{enerHW}\\
p_r^{eff}(r)&=&-\frac{6 m q^3}{\kappa ^2 \left(q^3+r^3\right)^2},\label{prHW}\\
p_t^{eff}(r)&=&-\frac{6 m q^3 \left(q^3-2 r^3\right)}{\kappa ^2 \left(q^3+r^3\right)^3}.\label{ptHW}
\end{eqnarray}
It's clear that we have the behavior of an anisotropic fluid with $\rho=-p_r$. In Fig. \ref{densHW} we can see how the fluid quantities, given by the equations \eqref{enerHW}-\eqref{ptHW}, behave in terms of the radial coordinate. The effective energy density is always positive and it's interesting to notice that all these quantities tend zero in the infinity and zero in the origin of the radial coordinate. It is also important to note that near the center of the black hole we have an isotropic behavior, $p_r\approx p_t$. The energy conditions are

\begin{eqnarray}
&&WEC_1(r)=0,\\
&&WEC_2(r)=\frac{18 m q^3 r^3}{\kappa ^2 \left(q^3+r^3\right)^3},\\
&&DEC_2(r)=2DEC_1(r)=\frac{12 m q^3}{\kappa ^2 \left(q^3+r^3\right)^2},\\
&&DEC_3(r)=\frac{6 m q^3 \left(2 q^3-r^3\right)}{\kappa ^2 \left(q^3+r^3\right)^3},\\
&&SEC(r)=-\frac{12 m q^3 \left(q^3-2 r^3\right)}{\kappa ^2 \left(q^3+r^3\right)^3}.
\end{eqnarray}
The energy conditions do not depend on the parameters of $f(G)$ gravity, being equal to the energy condition on general relativity in agreement with the theorem present in Sec. \ref{sec2}.
\subsection{Culetu solution}
\begin{figure*}
	\includegraphics[height=5.cm,width=7cm]{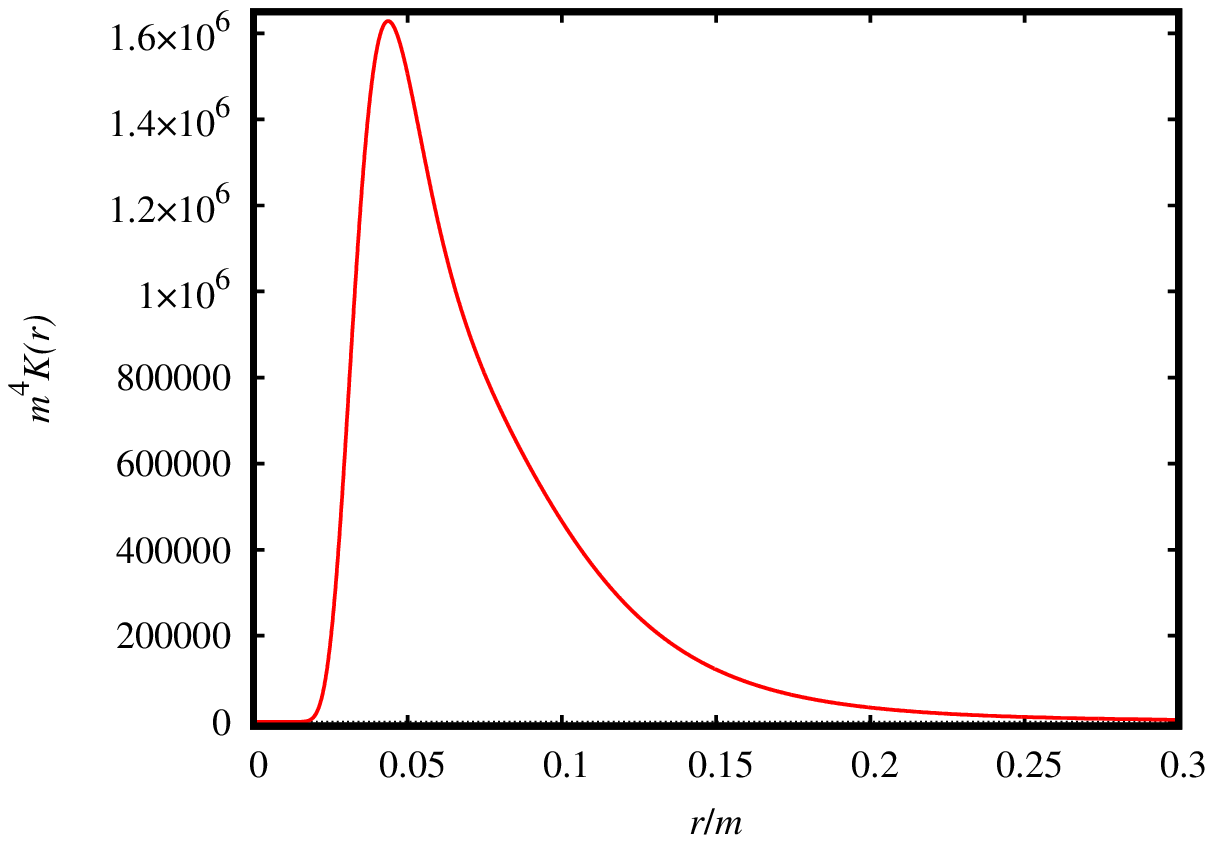}
	\includegraphics[height=5.cm,width=7cm]{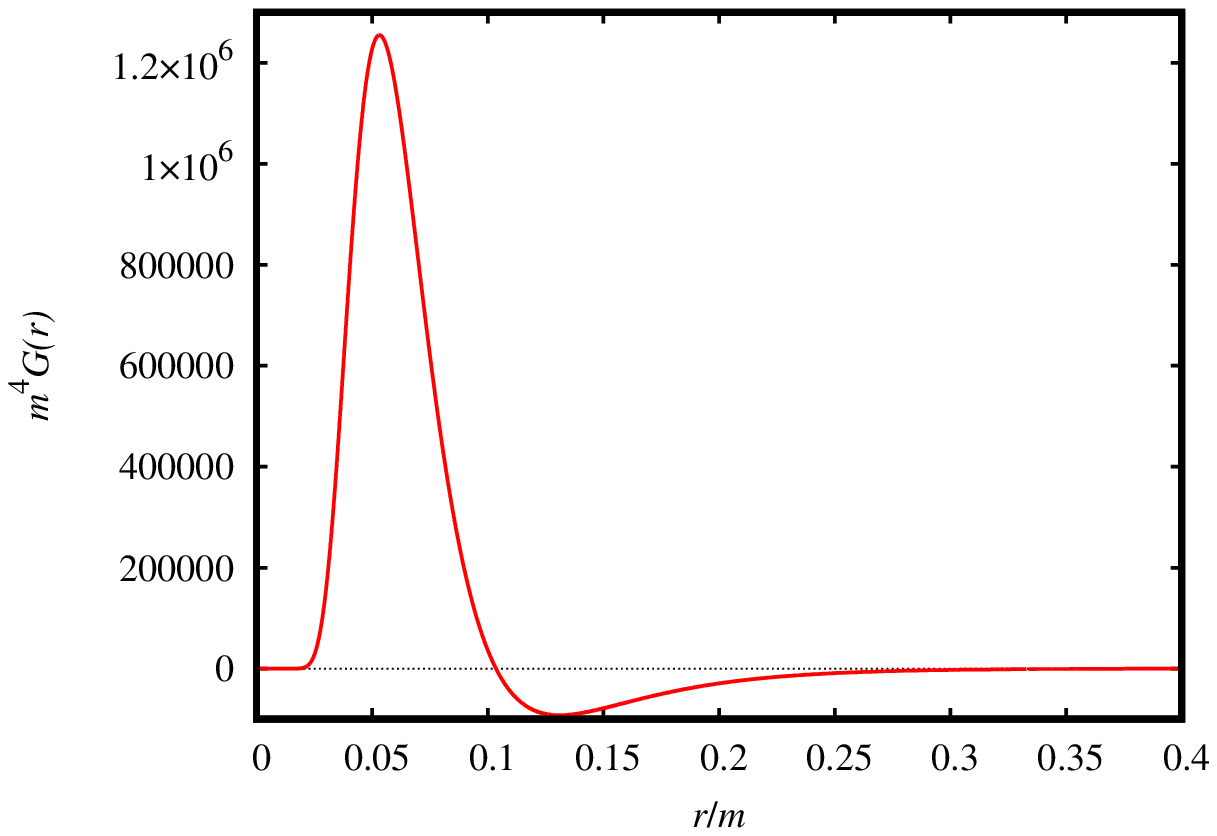}	
	\caption{Kretschmann scalar and Gauss-Bonnet invariant for the Culetu solution for $q=0.7m$.}
	\label{InvC}
\end{figure*}

\begin{figure*}
	\includegraphics[height=5.cm,width=7cm]{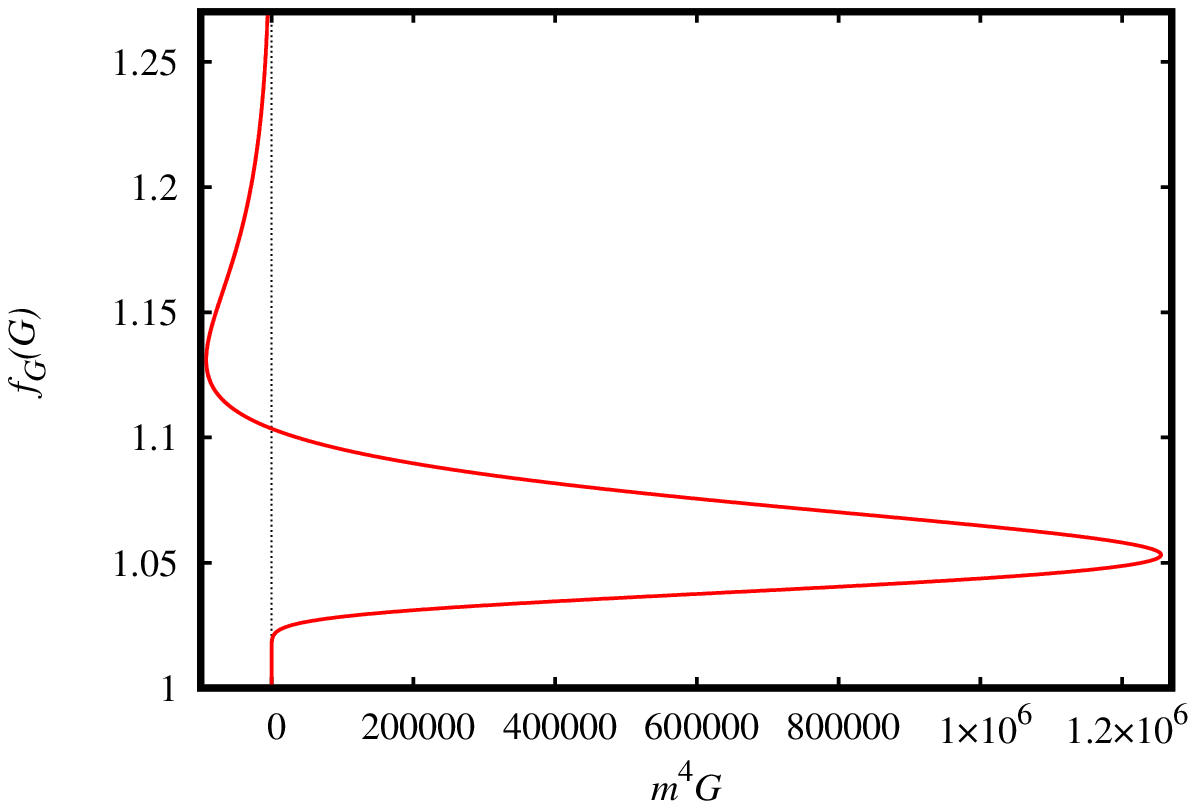}
	\includegraphics[height=5.cm,width=7cm]{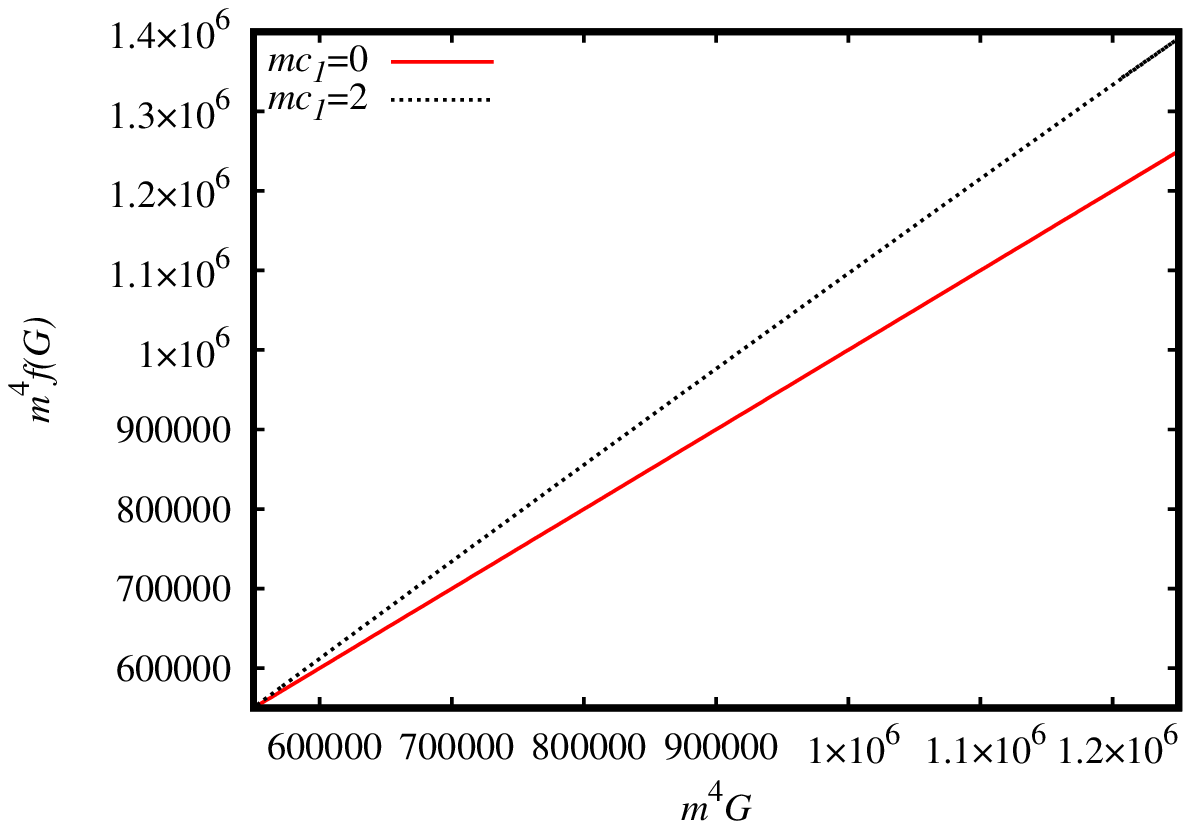}	
	\caption{Comparison between the $f(G)$ no the linear case, $mc_1=0$, and the nonlinear, $mc_1=2$, generates by the Culetu solution and the $f_G(G)$ function for $c_0=1$ and $q=0.7m$.}
	\label{fC}
\end{figure*}

In \cite{Culetu1,Culetu2} Culetu proposed a regular solution that is described by the mass function
\begin{equation}
M(r)=me^{-q^2/(2mr)}.
\end{equation}
This solution behaves asymptotically as Reissner-Nordström for regions far from the black hole and as de Sitter close to the center of the black hole. The Culetu solution has already been generalized for $f(R)$ gravity in \cite{rodrigues2}. From the curvature scalar, is possible construct an analytical expression for $r$ and then construct the $f(R)$ function to the Culetu solution. We will see that it is not so simple in $f(G)$ gravity. The Kretschmann scalar and the Gauss-Bonnet invariant are given by
\begin{eqnarray}
&&K(r)=\frac{e^{-\frac{q^2}{m r}}}{4 m^2 r^{10}} \left(192 m^4 r^4-192 m^3 q^2 r^3+96 m^2 q^4 r^2-16 m q^6 r+q^8\right),\\
&&G(r)=\frac{8 e^{-\frac{q^2}{m r}} \left(6 m^2 r^2-6 m q^2 r+q^4\right)}{r^8}.\label{GC}
\end{eqnarray}
The regularity of these functions is shown in the Fig. \ref{InvC}. We can see that the Kretschmann scalar zero far from the black hole and at the center and has a maximum. The Gauss-Bonnet has a minimum and a maximum value and goes to zero in the limits $r\rightarrow 0$ and $r\rightarrow \infty$. Different of the curvature scalar, it's not possible to invert $G(r)$ to obtain an analytical form of $r(G)$. So that, to show the nonlinear behave of the gravitational theory, we use \eqref{GC} with \eqref{fG} and $f(G)$ in terms of the radial coordinate, that is given by
\begin{eqnarray}
&&f(G)=\frac{e^{-\frac{q^2}{m r}}}{q^{10} r^8}, \left(8 c_0 q^{10} \left(6 m^2 r^2-6 m q^2 r+q^4\right)\right.-8 c_1 r \left(144 m^7 r^7+144 m^6 q^2 r^6+72 m^5 q^4 r^5+24 m^4 q^6 r^4\right.\nonumber\\
&&\left.\left.+6 m^3 q^8r^3-6 m^2 q^{10} r^2+7 m q^{12} r-q^{14}\right)\right),
\end{eqnarray}
to build a parametric plot, Fig. \ref{fC}, showing that the gravity theory is not the general relativity.

\begin{figure*}
	\includegraphics[height=5.cm,width=7cm]{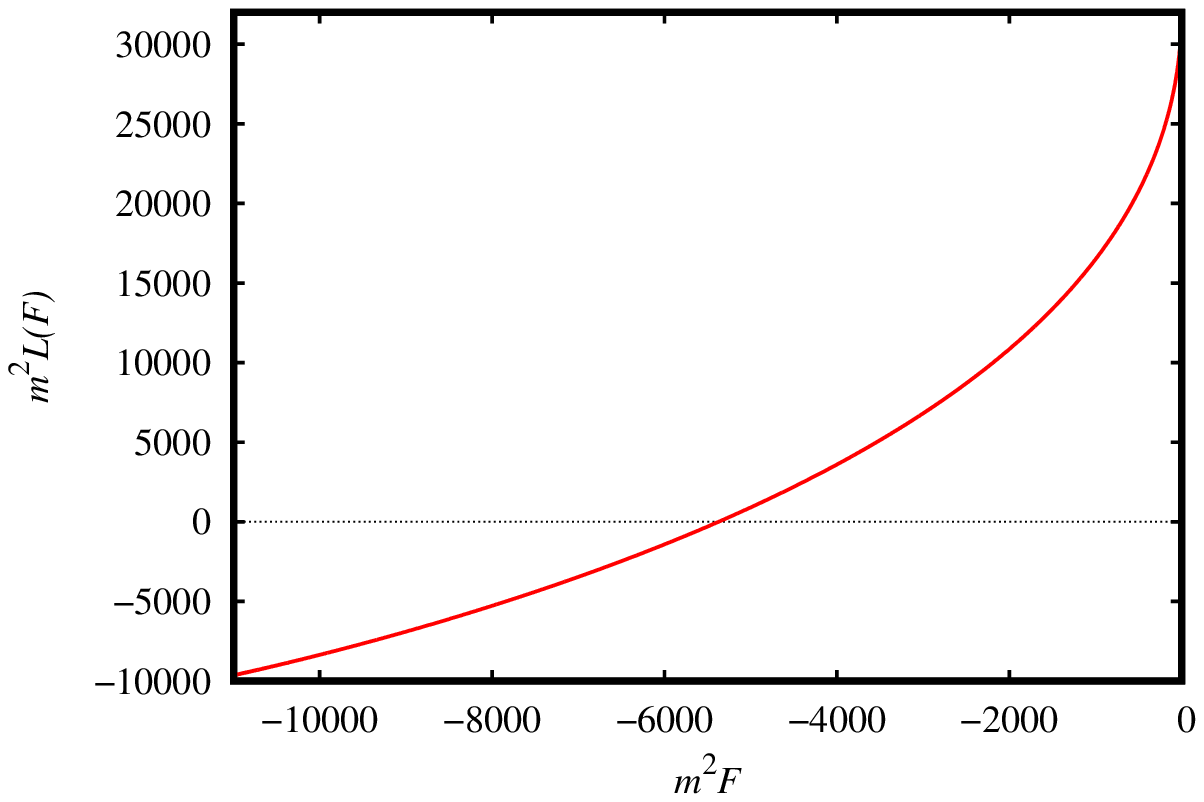}
	\includegraphics[height=5.cm,width=7cm]{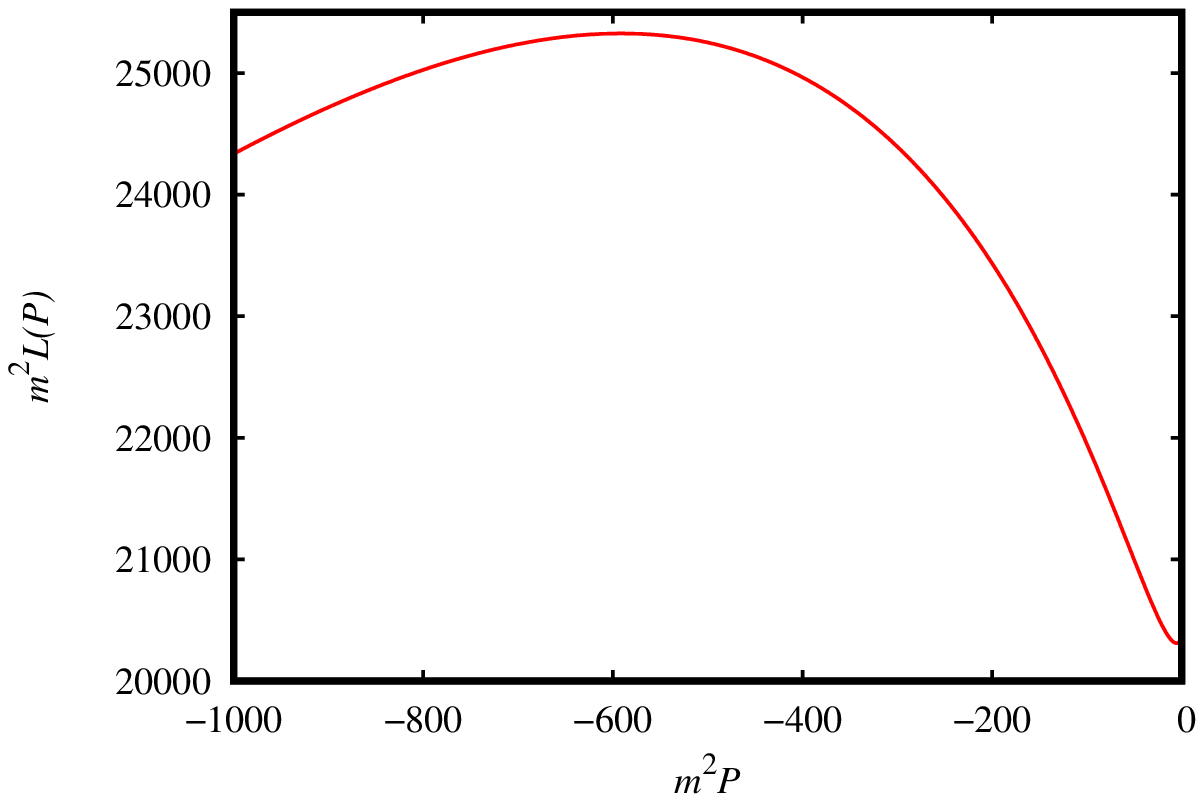}
	\caption{Graphical representation of the electromagnetic Lagrangian in terms of the electromagnetic scalar and the auxiliary field for $mc_1=1$ and $q=0.7m$.}\label{LC}
\end{figure*}

Substituting the mass function that generates the Culetu solution in \eqref{F10geral} and \eqref{Lgeral}, we find that the only nonzero component of the Faraday-Maxwell tensor and the electromagnetic Lagrangian, in terms of the radial coordinate, are 
\begin{eqnarray}
&&F^{10}=\frac{e^{-\frac{q^2}{m r}}}{8 m q r^5} \Big(32 c_1 m \left(12 m^2 r^2-9 m q^2 r+q^4\right)+r e^{\frac{q^2}{2 m r}} \big(8 m r^2 \left(q^2-16 c_1 m\right)-q^2 r \left(q^2-96c_1 m\right)8 c_1 q^4\big)\Big),\\
&&L=\frac{e^{-\frac{q^2}{m r}}}{8 r^7} \Bigg(\frac{r e^{\frac{q^2}{2 m r}}}{m} \left(8 c_1 \left(8 m^2 r^2-8 m q^2 r+q^4\right)+q^2 r \left(q^2-4 m r\right)\right)+\frac{32 c_1}{q^{10}}\left(144 m^7 r^7+144 m^6 q^2 r^6\right.\nonumber\\
&&+72 m^5 q^4 r^5+24 m^4 q^6 r^4+6 m^3 q^8 r^3-6 m^2 q^{10} r^2\left.+7 m q^{12} r-q^{14}\right)\Bigg).\label{LFC}
\end{eqnarray}
If we expand the electric field for larges values of the radial coordinate we find
\begin{equation}
F^{10}\approx \frac{4c_1}{qr^2} \left(\frac{12 m^2}{r}-4 m+\frac{5 q^2}{r}\right)-\frac{5 q^3}{8 m r^3}+\frac{q}{r^2}+O\left(\frac{1}{r^4}\right),
\end{equation}
which clearly is regular for $r\rightarrow \infty$. It's interesting since in the $f(R)$ gravity, due to the coupled with the gravitational theory, the electric field diverges for $r\rightarrow \infty$ (see equation A5 in \cite{rodrigues2}). From \eqref{LFC}, we show the nonlinear behavior of the electromagnetic theory in Fig. \ref{LC} where we analyze $L$ in terms of the scalars $F$ and $P$.  The analytical expression for $L(P)$ is given by \eqref{LPC}.

At least, in order to find the energy conditions, we calculate the fluid quantities, that are
\begin{eqnarray}
&&\rho^{eff}(r)=\frac{q^2 e^{-\frac{q^2}{2 m r}}}{\kappa ^2 r^4},\\
&&p_r^{eff}(r)=-\frac{q^2 e^{-\frac{q^2}{2 m r}}}{\kappa ^2 r^4},\\
&&p_t^{eff}(r)=\frac{q^2 e^{-\frac{q^2}{2 m r}} \left(4 m r-q^2\right)}{4 \kappa ^2 m r^5}.
\end{eqnarray}
With that, the energy conditions become
\begin{eqnarray}
&&WEC_1(r)=0,\\
&&WEC_2(r)=\frac{q^2 e^{-\frac{q^2}{2 m r}} \left(8 m r-q^2\right)}{4 \kappa ^2 m r^5},\\
&&DEC_2(r)=2DEC_1(r)=\frac{2 q^2 e^{-\frac{q^2}{2 m r}}}{\kappa ^2 r^4},\\
&&DEC_3(r)=\frac{q^4 e^{-\frac{q^2}{2 m r}}}{4 \kappa ^2 m r^5},\\
&&SEC(r)=\frac{q^2 e^{-\frac{q^2}{2 m r}} \left(4 m r-q^2\right)}{2 \kappa ^2 m r^5}.
\end{eqnarray}
So, as in general relativity and in $f(R)$ gravity, $SEC$ and $NEC_2$ are violated inside the event horizon while the other conditions are always satisfied.
\section{Conclusion}\label{sec5}
In this work, we have developed a method that generalizes solutions of regular black holes, already known from general relativity, to the $f(G)$ theory. The method consists of writing the gravitational and electromagnetic quantities in terms of a mass function, in such a way that each solution generates a different electromagnetic and gravitational theory. Through the equations of motion and considering the symmetry $a=-b$, we show that the $f_G(G)$ function has a linear dependence on the radial coordinate, which clearly diverge in the limit $r\rightarrow \infty$. The divergence in $f_G(G)$ does not imply that there will also be divergences in $f(G)$. For the models of regular black hole present here $f(G)$ is always regular.

As examples of black holes, we construct the generalization of the Schwarzschild and RNAdS solutions and we showed that in the $f(G)$ gravity the Schwarzschild solution could not be interpreted as a vacuum solution. In the Schwarzschild case, we obtained an analytical form for $f(G)$ and a numerical form to RNAdS. The linear term with the Gauss invariant usually does not modify the equations of motion, however, for RNAdS this term is coupled with the cosmological constant such that to recover the results of general relativity it is necessary to make both $c_1=0$ and $c_0 =0$, which is when we recover the linearity of the electromagnetic theory.

For the regular models, we choose a mass function that has the Bardeen and the Hayward solution as specific cases. We also chose the mass function that generates the Culetu solution, so that we can compare this result with those already known from general relativity and the $f(R)$ theory. For the case of coupling with general relativity, the electric field associated with the source is always regular, tends to zero at the origin and at infinity and has a maximum point. For the $f(R)$ gravity, the electric field diverge in infinity and therefore it is necessary to analyze the electric induction tensor. This type of divergence does not appear in the case of coupling with the $f(G)$ theory, but now the electric field has maximum and minimum points.

Since it is not possible to find an analytical form to $L(F)$, we obtain the nonlinear behavior of these functions numerically. As we are working with electrical sources, it becomes useful to use the auxiliary field $P$ and with it, we can find a closed form for $L(P)$.

The energy conditions are the same as those obtained in general relativity and in the $f(R)$. For all solutions, the strong energy condition is violated inside the black hole, which implies that within the event horizon we have regions in which the gravitational interaction is repulsive. For the Culetu solution, we have that the null energy condition is violated outside the event horizon and for the other solutions the dominant energy condition is violated outside the black hole.

\textbf{As continuations of this work, we can try to verify the stability and the possibility of application of the $f(G)$ models developed here to compare with cosmological data \cite{felice2}. To do that, since in the Newtonian limit we do not have corrections from the perturbations in $f(G)$ gravity in relation to the general relativity (compare (22) for $f(R,G)=R$ with (51) in \cite{maria}), it's necessary to use the Post Newtonian and Parameterized Post Newtonian limit to get new corrections. With that, it's possible to find a regime of validity for $c_1$, as was done in \cite{felice3} using the deflection of light, Cassini experiment, perihelion shift retardation of light, gravitational redshift and the equivalence principle.}
\vspace{1cm}

{\bf Acknowledgements}: M. E. R.  thanks Conselho Nacional de Desenvolvimento Cient\'ifico e Tecnol\'ogico - CNPq, Brazil, Edital MCTI/CNPQ/Universal 14/2014  for partial financial support.

\appendix
\section{Analytical forms}\label{AF}
The analytical forms for some functions are too much large. So that, we dedicate this appendix to show the analytical expressions.
\subsection{Reissner-Nordström-anti-de Sitter}
Due to the coupled with $f(G)$ gravity and the presence of the cosmological constant, the electromagnetic theory that generates RNAdS is not linear, $L(P)=P$, actually we have

\begin{eqnarray}
&&L(P)=P+\frac{4 c_0 \Lambda ^2}{3}+\frac{4 c_1 P\sqrt[4]{-P}}{105 q^{5/2}} \left[1512 \sqrt[4]{2} m^2 P+420 m \sqrt[4]{-q^2P} \left(\sqrt{-P}-7 \sqrt{2} P q\right)\right.\nonumber\\
&&\left.+5 \sqrt[4]{2} q^2 \left(6 P\left(40 \sqrt{-2P} q+21\right)-7 \Lambda \right)\right].
\end{eqnarray} 
So that, the generalized RNAdS is characterized by a nonlinear electrodynamics. If we are free of the cosmological constant, $c_1=0$ is enough to recover the Maxwell theory and the constant $c_0$ doesn't appear in the electromagnetic functions.

\subsection{First regular solution}
To regular black holes with electric sources is not possible write an analytical form to $L(F)$, however is still possible find to $L(P)$. Since we have $r(P)$ and $r(G)$, we replaced $r(P)$ in \eqref{L0BDG} and $r(G)$ in \eqref{fG} and \eqref{fGBDG} to find the analytical expressions. The electromagnetic Lagrangian is

\begin{eqnarray}
&&L(P)=c_1 \left\{\frac{2 m^2}{q^5} \left[\sqrt{3} \ln\left(\frac{q+ \sqrt[4]{-18q^6P}+q^2\sqrt{-2P}}{q-\sqrt[4]{-18q^6P}+q^2\sqrt{-2P}}\right)\right.\right.+2 \tan ^{-1}\left(\sqrt[4]{\frac{-8}{q^2P}}+\sqrt{3}\right)\nonumber\\
&&-2 \tan ^{-1}\left(\sqrt[4]{\frac{-8}{q^2P^5}}P+\sqrt{3}\right)\left.+4 \tan ^{-1}\left(\frac{1}{\sqrt[4]{-2q^2P}}\right)\right]+\frac{8 m \sqrt[4]{-2P} P}{q^{7/2} \left(8 P^3 q^6+1\right)^3}\Bigg[6 m q \left(\sqrt{\frac{\sqrt{2} q^3}{(-P)^{3/2}}+4 q^6}\times\right.\nonumber\\
&&\left( 4 P^2 q^6 \left(77 \sqrt{-2P}-308 P^2 q^3+64 P^5 q^9+232 \sqrt{2} (-P)^{7/2} q^6\right)+58 P^2q^3 -\sqrt{-2P}\right)+16 P^3 q^6 \left(52 P^3 q^6\right.\nonumber\\
&&\left.+8 \sqrt{2} (-q^2P)^{9/2}+34 (-8q^2P)^{3/2}-17\right)\left.\left.+26 \sqrt{-2P} P q^3+1\right)\Bigg]\right\}+\frac{6 m P^5 \sqrt{\frac{\sqrt{2} q^3}{(-P)^{3/2}}+4 q^6}}{\left(4 \sqrt{2} P^2 q^3+8 (-P)^{7/2} q^6+\sqrt{-P}\right)^4}\nonumber\\
&&\times \left(32 P^3 q^6 \left(32 P^6 q^{12}+30 P^3 q^6+12 \sqrt{2} (-P)^{9/2} q^9+25 \sqrt{-2P} P q^3-15\right)+66 \sqrt{2} (-q^2P)^{3/2}+7\right).
\label{LPBDG}
\end{eqnarray}
In the weak field limit, we do not recover the Maxwell theory since, for $c_1=0$, we have $L(P)\approx P^{9/4}$ with $P\ll 1$. 

Using \eqref{GBDG} we get $r$ as a function of $G$ and substituting \eqref{fG} we get
\begin{eqnarray}
&&f_G(G)=c_0+\frac{c_1}{G^{1/6}}\Bigg\{m^2 \left(16-\frac{44 \sqrt[3]{2} G q^6}{\alpha_1(G)}\right)+2\ 2^{2/3} \alpha_1(G)+\frac{64 \sqrt[3]{2} m^4}{\alpha_1(G)}-G q^6\Bigg\}^{1/6}.\label{FGGAN1}
\end{eqnarray}
Where $\alpha_1(G)$ is given by
\begin{eqnarray}
&&\alpha_1(G)=\Big\{9 G^2 m^2 q^{12}-132 G m^4q^6+128 m^6+\sqrt{G^2 m^4 q^{12} \left(81 G^2 q^{12}+2948 G m^2 q^6-3504 m^4\right)}\Big\}^{1/3}\nonumber.
\end{eqnarray}

Integrating \eqref{FGGAN1} with respect to $G$ and defining
\begin{equation}
\alpha_2(G)=\frac{1}{G}\Bigg[4 m^2 \left(4-\frac{11 \sqrt[3]{2} G q^6}{\alpha_1(G)}\right)+2^{5/3} \alpha_1(G)-G q^6+\frac{64 \sqrt[3]{2} m^4}{\alpha_1(G)}\Bigg],\nonumber
\end{equation}
we obtain
\begin{eqnarray}
&&f(G)=c_0 G+4 c_1 m^2\left(\frac{1}{q^5}\left[4 \cot ^{-1}\left(q\alpha_2^{-1/6}(G)\right)\right.\right.-2 \tan ^{-1}\left(-\frac{2}{q}\alpha_2^{1/6}(G)+\sqrt{3}\right)+2 \tan ^{-1}\Bigg(\frac{2}{q}\times \nonumber\\
&&\left.\alpha_2^{1/6}(G)+\sqrt{3}\Bigg)+\sqrt{3} \Big(\ln \Big(q^2+\sqrt{3}\alpha_2^{1/6}(G) q+\alpha_2^{1/3}(G)\Big)-\ln \Big(q^2-\sqrt{3} \alpha_2^{1/6}(G) q+\alpha_2^{1/3}(G)\Big)\Big)\right]\nonumber\\
&&\left.-\frac{3 G^3 \alpha_1^3(G) \alpha_2^{1/6}(G)\left(q^{12}-10q^6 \alpha_2(G) +\alpha_2^2(G)\right)}{2 \left(32 \sqrt[3]{2} m^4+\left(8 \alpha_1(G)-22 \sqrt[3]{2} G q^6\right) m^2+2^{2/3} \alpha_1^2(G)\right)^3} \right).\label{FGAN1}
\end{eqnarray}

\subsection{Hayward-type solution}
From \eqref{gHW}, we may construct an analytical expression to $r(G)$. Defining
\begin{eqnarray}
&&\alpha_3(G)=3 \sqrt{G m^6 q^6 \left(81 G^2 q^{12}+792 G m^2 q^6-112 m^4\right)}+90 G m^4 q^6-8 m^6,\\
&&\alpha_4(G)=\frac{1}{\sqrt{G \sqrt[3]{\alpha_3(G)}}}\bigg(4 m^4+\left(4 \sqrt[3]{\alpha_3(G)}-9 G q^6\right) m^2+\left(\alpha_3(G)\right)^{2/3}\bigg)^{1/2},\\
&&\alpha_5(G)=\frac{1}{\sqrt{G}}\bigg(\frac{24 \sqrt{2} m^2 q^3}{\alpha_4(G)}+8 m^2-\sqrt[3]{\alpha_3(G)}+\frac{9 G m^2 q^6-4 m^4}{\sqrt[3]{\alpha_3(G)}}\bigg)^{1/2},
\end{eqnarray}
we get the $f_G(G)$ function, that is given by 
\begin{eqnarray}
f_G(G)=c_0-c_1 \left(\sqrt{2} \alpha_4(G)+q^3-\sqrt{2}\alpha_5\right).\label{anafGHW}
\end{eqnarray}
Integrating this equation with respect to $G$ we get the analytical form of $f(G)$. For
\begin{eqnarray}
&&\alpha_6(G)= \sqrt[3]{-q^3-\sqrt{2}\alpha_4(G) +\sqrt{2} \alpha_5(G)},
\end{eqnarray}
$f(G)$ is written as
\begin{eqnarray}
&&f(G)=c_0 G-\frac{16 c_1 m^2}{3 q^5} \Bigg(\frac{3\alpha_6(G) \left(-2 q^9+24 \alpha^3_6(G)
q^6+\alpha^9_6(G)\right) q^2}{4 \left(\alpha_4(G)-\alpha_5(G)\right)^4}+2 \sqrt{3} \tan ^{-1}\left(\frac{2 \alpha_6(G)-q}{\sqrt{3} q}\right)\nonumber\\
&&+2 \ln \left(q+\alpha_6(G)\right)-\ln \left(q^2-\alpha_6(G)q+\alpha^2_6(G)\right)\Bigg).
\label{anafHW}
\end{eqnarray}

In terms of the auxiliary field $P$, the electromagnetic Lagrangian that generates the Hayward-type solution in $f(G)$ gravity, \eqref{L0HW}, is
\begin{eqnarray}
&&L(P)=\frac{8 c_1 m }{3
	q^5 \left(2 (-q^2P)^{3/4} +\sqrt[4]{2}\right)^5}\left(2 m \left(40 \sqrt[4]{2} P^3 q^6-16 (-q^2P)^{15/4}-40 \sqrt{2} (-q^2P)^{9/4}-10 (-Pq^2)^{3/4}\right.\right.\nonumber\\
&&\left.+20\ 2^{3/4} \sqrt{-P} P q^3-\sqrt[4]{2}\right)
	\left(\ln \left(\frac{q- \sqrt[4]{-2P} q^{3/2}+q^2 \sqrt{-2P}}{q+2 \sqrt[4]{-2P} q^{3/2}+q^2 \sqrt{-2P}}\right)+2 \sqrt{3} \tan	^{-1}\left(\frac{2^{3/4} P}{(-P)^{5/4} \sqrt{3q}}+\frac{1}{\sqrt{3}}\right)\right)\nonumber\\
&&+6 m q \left(96 \sqrt[4]{2} P^2 q^3+176 (-P)^{11/4} q^{9/2}+2 \sqrt{2} (-P)^{5/4} q^{3/2}-16\ 2^{3/4}(-P)^{7/2} q^6+2^{3/4} \sqrt{-P}\right)\nonumber\\
&&+12 P q^3 \left(8 \sqrt[4]{2} P^3 q^6\left.+20 \sqrt{2} (-q^2P)^{9/4}+10 (-q^2P)^{3/4}-\sqrt[4]{2}+24\ 2^{3/4} (-q^2P)^{3/2}\right)\right)\nonumber\\
&&-\frac{12\sqrt{2} m P\left(\sqrt{-2q^2P}- \sqrt[4]{-2q^2P}+1\right)^{-2}}{\left( \sqrt[4]{-2q^2P} +2\right)  \left((-2q^2P)^{3/4}+1\right)^5}\left(8 P^4 q^7+4 P^3 q^5-6 P^2 q^3\right.-4\ 2^{3/4} (-P)^{15/4} q^{13/2}\nonumber\\
&&+4 \sqrt[4]{2} (-P)^{13/4} q^{11/2}+2\ 2^{3/4}(-P)^{11/4} q^{9/2}+6 \sqrt[4]{2} (-P)^{9/4} q^{7/2}+5(-2P)^{3/4} P q^{5/2}+5 \sqrt[4]{-2P} P q^{3/2}\nonumber\\
&&+4 \sqrt{2} (-P)^{7/2} q^6-6 \sqrt{2} (-P)^{5/2} q^4+5 \sqrt{2}(-P)^{3/2} q^2+2 P q+ (-2P)^{3/4} \sqrt{q}\left.-\sqrt{-2P}\right).
\label{LPHW}
\end{eqnarray}

\subsection{Culetu solution}
For the Culetu solution is not possible write $f_G(G)$ and $f(G)$ in a closed. However, it is still possible to construct for $L(P)$, that is

\begin{eqnarray}
&&L(P)=\frac{2 c_1 e^{-\frac{\sqrt[4]{-2q^6P}}{m}} }{m q^{10}}\left(288 m^8+288  m^7 \sqrt[4]{-2q^6P}+144  m^6 \sqrt{-2q^6P}+48\ m^5 (-2q^6P)^{3/4}
-24 m^4 P q^6\right.\nonumber\\
&&.+24  m^3 \sqrt[4]{-2q^{30}P} P-4 m^2 P q^8 \left(2 e^{\frac{\sqrt[4]{-q^6P/8}}{m}}+7 \sqrt{-2q^2P}\right)+4  m \sqrt[4]{-2P}
P q^{19/2} \left(2 e^{\frac{\sqrt[4]{-q^6P/8}}{ m}}\right.\nonumber\\
&&\left.\left.+\sqrt{-2q^2P} \right)+(-8P)^{3/2} q^{11} e^{\frac{\sqrt[4]{-q^6P/8}}{ m}}\right)+\frac{Pe^{-\frac{\sqrt[4]{-q^6P/8} }{ m}} \left(4 m (-P)^{3/4}+\sqrt[4]{2q^6} P \right)-\frac{4}{q^{10}}}{4 m (-P)^{3/4}}.\label{LPC}
\end{eqnarray}

\end{document}